\documentclass[%
 reprint,
superscriptaddress,
 amsmath,amssymb,
 aps,
]{revtex4-2}

\usepackage{graphicx}
\usepackage{dcolumn}
\usepackage{bm}
\usepackage{bbm}
\usepackage{subcaption}
\usepackage{comment}
\usepackage{physics}
\usepackage{amsmath}
\usepackage{bbold}
\usepackage{caption}
\usepackage{xcolor}

\begin{document}

\preprint{APS/123-QED}

\title{Highly Squeezed States in Ring Resonators:\\
Beyond the Undepleted Pump Approximation}
\author{Colin Vendromin}
\email{colin.vendromin@utoronto.ca}
\affiliation{Department of Physics, University of Toronto, 60 St. George Street, Toronto, Ontario, Canada M5S 1A7}%
\author{Yan Liu}
\affiliation{School of Physics Science and Information Technology, Shandong Key Laboratory of Optical Communication Science and Technology, Liaocheng University, Liaocheng 252059, China}
\author{Zhenshan Yang}
\affiliation{School of Physics Science and Information Technology, Shandong Key Laboratory of Optical Communication Science and Technology, Liaocheng University, Liaocheng 252059, China}
\author{J. E. Sipe}
\affiliation{Department of Physics, University of Toronto, 60 St. George Street, Toronto, Ontario, Canada M5S 1A7}%

\date{\today}
\begin{abstract}
We present a multimode theory of squeezed state generation in resonant systems valid for arbitrary pump power and including pump depletion. The Hamiltonian is written in terms of asymptotic-in and -out fields from scattering theory, capable of describing a general interaction.  As an example we consider the lossy generation of a highly squeezed state by an effective second-order interaction in a silicon nitride ring resonator point-coupled to a waveguide. We calculate the photon number, Schmidt number, and the second-order correlation function  of the generated state in the waveguide. The treatment we present provides a path forward to study the deterministic generation of non-Gaussian states in resonant systems.
\end{abstract}

\maketitle
\section{Introduction}
Continuous-variable (CV) squeezed states of light  exhibit an uncertainty below vacuum noise in one quadrature \cite{schnabelSqueezeStates} making them an essential resource for implementing CV quantum information processing \cite{Weedbrook2012}, and  with post-selection strategies they can be used to create
non-Gaussian states of light \cite{Wenger2004, Endo2023}, an important resource for quantum computing. 

Squeezed states are typically generated through nonlinear optical processes such as spontaneous parametric down conversion (SPDC) and spontaneous four-wave mixing (SFWM), the first relying on the presence of a $\chi^{(2)}$ susceptibility, and the second on the $\chi^{(3)}$ susceptibility \cite{Slusher1987, Zhao2020}.  In the usual treatment of these processes, the pump field or fields are treated classically, with the nonlinear process being considered sufficiently weak that those fields can be treated as undepleted. 
%
In this case the resulting squeezed state can be approximately written as \cite{backwardHeisenberg}
\begin{align}
    \label{eq:intro - BHP ket}
    \ket{\psi}= \exp(\frac{\beta}{2} \int dk_1 k_2 \phi(k_1, k_2) a^\dagger_{k_1}a^\dagger_{k_2} -{\rm H.c.})\ket{vac},
\end{align}
where  $\beta$ is an overall squeezing parameter, and $\phi(k_1, k_2)$ is the normalized biphoton wavefunction that describes the production of pairs of photons; one with wavenumber $k_1$ and the other $k_2$. The ket in Eq. \eqref{eq:intro - BHP ket} is equivalent to a lowest-order Magnus expansion \cite{Blanes2009} that treats the pump as classical and  undepleted. If the squeezing parameter $\beta$ is sufficiently weak, then time-ordering corrections that arise in the higher-order terms of the Magnus expansion are expected to be negligible. However, time-ordering effects  become important when the amount of squeezing increases \cite{Triginer2020, beyondPhotonPairs, quesada2014}, and the approximate ket in Eq. \eqref{eq:intro - BHP ket} then becomes a poorer and poorer approximation.

Including time-ordering corrections, the ket can obtained
under the assumption that the pump is classical and undepleted,  and that the generated light is Gaussian. As a result, the full Gaussian ket can be written generally as a squeezed vacuum state
\begin{align}
    \label{eq:intro - full ket}
    \ket{\psi} = \exp(\frac{1}{2}\int dk_1 dk_2 J(k_1,k_2)a^\dagger_{k_1}a^\dagger_{k_2} - {\rm H.c.})\ket{vac},
\end{align}
where the squeezing matrix $J(k_1, k_2)$ can be obtained from the Heisenberg equations of motion \cite{beyondPhotonPairs}. If time-ordering corrections are negligible this ket reduces to  Eq. \eqref{eq:intro - BHP ket}, with  the squeezing matrix just being proportional to the biphoton wavefuction  $J(k_1, k_2) \approx \beta \phi(k_1, k_2)$.

However, if the nonlinear process is strong enough that pump depletion needs to be considered, the generated light can be expected to be non-Gaussian due to entanglement between the pump and the generated fields \cite{Yanagimoto2022, quesada2023}. In this case a squeezed vacuum state such as in Eq. \eqref{eq:intro - full ket} does not provide a complete description of the features of the generated light, such as pump depletion. To go beyond the undepleted pump approximation  one strategy that has been employed is to replace the vacuum state in Eq. \eqref{eq:intro - full ket} with a non-Gaussian ket, which starts from vacuum initially and evolves such that it remains close to the vacuum. This treatment was recently investigated in waveguides \cite{Yanagimoto2022} and for  a single-mode SPDC Hamiltonian \cite{quesada2023}, and offers a potential route to the deterministic generation of non-Gaussian states of light. In fact experiments have already demonstrated high conversion efficiencies where the undepleted pump approximation is no longer valid \cite{Florez2020,ramelow2019}, and non-Gaussianity may become important.

In this paper  we extend such a treatment to   resonant systems. 
As an example we study a ring resonator, a system of interest in its own right due to its resonant enhancement of nonlinear processes and the precise control over the frequencies involved, as well as its integrability on photonic chips \cite{Steiner2021, Vaidya2020, Zhang2021}. We develop a multimode Hamiltonian theory of non-Gaussian state generation in such systems, including scattering loss, by employing
the asymptotic-in and -out fields used in scattering theory \cite{Liscidni2012}. 

We apply the theory to the particular case of generating a highly squeezed state by an effective $\chi^{(2)}$ interaction in a ring resonator \cite{xanaduCV2019,ramelow2019}. For this interaction one relies on the $\chi^{(3)}$ response of the medium, but there is one pump field in a mode $C$ that is continuous wave (CW) and so strong that it can be treated as classical and undepleted; in the presence of the material $\chi^{(3)}$ this leads to an effective $\chi^{(2)}$ through which a much weaker pump in the mode $P$ can generate two photons in  the mode $S$ (see Fig. \ref{fig:ring}(a)).  There is no issue with phase-matching, since the pumps and the generated photons are close in wavelength. The formalism we present can also be applied to a standard $\chi^{(2)}$ interaction as well.

Although the theory we present is capable of studying non-Gaussian states, our focus in this paper is on highly squeezed states that include the effects of pump depletion. The  treatment of the non-Gaussian corrections to the squeezed state will be handled in a subsequent communication, as will the effects of self- and cross-phase modulation, and the excitation of fields in additional ring resonances. While in this paper we specialize to ring resonator systems, most of the formalism can be applied to other systems as well, such as waveguides.

This paper is organized as follows. In Sec. \ref{sec:H effective chi2} we write
the system Hamiltonian, including the SFWM interaction in a ring resonator, in terms of the asymptotic-in and -out fields for our system. In Sec. \ref{sec:inoutstates} we introduce the input and output kets  that are used to describe the generation of light in the ring resonator system. In Sec. \ref{sec:Ubar} we derive the non-Gaussian unitary operator that connects the input ket to the output ket. In Sec. \ref{sec:low} we show how to obtain the first-order solution for the ket. In Sec. \ref{sec:numericalring} we numerically study the generation of a highly squeezed state by an effective $\chi^{(2)}$ interaction in a ring resonator with loss. In Sec. \ref{sec:conclude} we conclude.

\section{The Hamiltonian}
\label{sec:H effective chi2}
\begin{figure}[htbp]
    \centering
    \includegraphics[scale=0.25]{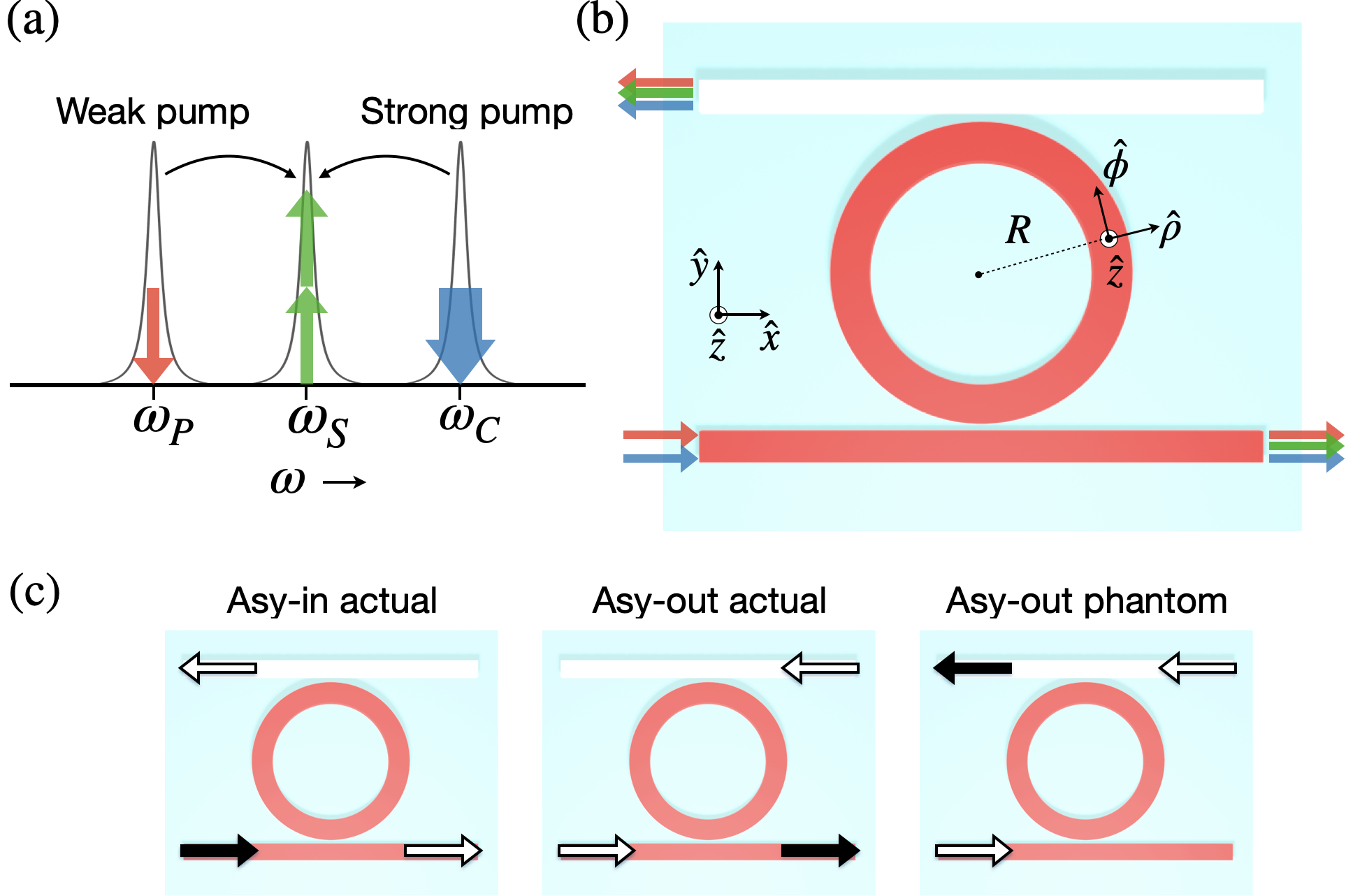}
    \caption{\small (a) SFWM with a weak pump at $\omega_P$ and a strong pump at $\omega_C$. (b) Schematic of a ring resonator point-coupled to an actual waveguide (bottom red) and phantom waveguide (top white). (c) Asymptotic-in field associated with the input channel of the actual waveguide (left), and asymptotic-out fields associated with the  output channel of the actual waveguide (center) and  phantom waveguide (right), respectively \cite{Liscidni2012,MBanicRingLoss2022}.}
    \label{fig:ring}
\end{figure}

Our system of interest is a ring resonator point-coupled to an actual waveguide and a phantom waveguide (see Fig. \ref{fig:ring}(b)), where the phantom waveguide is used to describe scattering losses \cite{MBanicRingLoss2022}. We define the ``input and  output channels" of the actual waveguide (shown in red) as the parts of that 
waveguide that are to the left and to the right of the coupling point, respectively; likewise, the input and output channels of the phantom waveguide (shown in white) are the parts of that  waveguide  that are
to the right and to the left of the coupling point, respectively.

As a basis for expanding the electromagnetic field in the full system of waveguides and ring, we employ the asymptotic-in and -out mode fields \cite{Liscidni2012, MBanicRingLoss2022}; they are sketched in Fig. \ref{fig:ring}(c), and are eigenstates of the full linear Hamiltonian of the system. Each asymptotic-in or -out mode field has an amplitude in both waveguides and in the ring. In each diagram, the black arrow corresponds to a freely propagating field that is either incoming (asymptotic-in) or outgoing (asymptotic-out), with its amplitude equal to that of a mode field in an isolated waveguide. The white arrows indicate either outgoing (asymptotic-in) or incoming (asymptotic-out) fields in the other channels, and their amplitudes depend on the coupling parameters between the ring and the waveguides \cite{MBanicRingLoss2022}. Asymptotic-in mode fields are also defined for the phantom channel, but they will not appear in our calculations.

The asymptotic-in and -out mode fields can be identified by three parameters. We use a discrete index $n$ to indicate the actual output channel or the phantom output channel, the waveguide wavenumbers
$k$, and an index $J$ that identifies the range of wavenumbers $k$ in the  
waveguide that correspond to a range of frequencies near a ring resonance, 
where $J=P,\, S,\, C$. We assume the ring resonances have
sufficiently high quality factors that the ranges over which $k$ varies for each $J$ can be considered distinct; we leave this implicit in the integrals below.

For the system considered here, the displacement field operator 
can be written in terms of the asymptotic-in mode fields  as
\cite{Liscidni2012}
\begin{align}
\label{eq:Din}
\bm D(\bm r) &= \sum_{n, J} \int dk  b_{n J k} \bm D^{\text{\textit{in}}}_{n J k}(\bm r) + {\rm H.c.},
\end{align}
where $b_{nJk}$ is the operator that annihilates an asymptotic-in photon associated with input channel $n$, with wavenumber $k$ in the ring resonance  $J$, and $ \bm D^{\text{\textit{in}}}_{n J k}(\bm r)$ is the mode field associated with that photon (see the left sketch in Fig. \ref{fig:ring}(c) for the asymptotic-in mode field associated with the actual input channel). Similarly, the displacement field operator can be written in terms of the asymptotic-out mode fields as 
\begin{align}
\label{eq:Dout}
\bm D(\bm r) &= \sum_{n, J} \int dk a_{n J k} \bm D^{\text{\textit{out}}}_{n J k}(\bm r) + {\rm H.c.},
\end{align}
where $a_{nJk}$ is the operator that annihilates an asymptotic-out photon associated with output channel $n$, with wavenumber $k$ in the ring resonance $J$, and $\bm D^{\text{\textit{out}}}_{n J k}(\bm r)$ is the mode field associated with that photon (see the middle or right sketch in Fig. \ref{fig:ring}(c)). The input and output operators satisfy the commutation relations 
\begin{align}
 \label{eq:commute}
 [a_{n J k}, a^\dagger_{n J k'}] = [b_{n J k}, b^\dagger_{n J k'}] = \delta(k-k'), 
 \end{align}
 with all others zero; the linear Hamiltonian can be written in terms of the input or output operators as
\begin{align}
\label{eq:HL}
H_{L} &= \sum_{n,J}\int dk \hbar \omega_{n J k}b^\dagger_{n J k} b_{n J k} \nonumber
\\
&= \sum_{n, J}\int dk \hbar \omega_{n J k}a^\dagger_{n J k} a_{n J k}, 
\end{align}
where $\omega_{nJk}$ is the dispersion relation for the waveguide containing channel $n$, ranging only over wavenumbers $k$ near ring resonance $J$; see Appendix \ref{sec:ring example} for details.


We now include a third-order nonlinear interaction described by
the Hamiltonian \cite{Quesada2022}
\begin{align}
    \label{eq:HNL}
    H_{\textit{NL}} = -\frac{1}{4\epsilon_0}\int d\bm r \Gamma^{ijkl}_3(\bm r) D^i(\bm r) D^j(\bm r)D^k(\bm r)D^l(\bm r),
\end{align}
where $i$, $j$, $k$, and $l$ are Cartesian components, as usual summed over when repeated, and the tensor $\Gamma^{ijkl}_3(\bm r)$ is related to the more familiar third-order tensor $\chi^{ijkl}_3(\bm r)$ by \cite{Quesada2022}\begin{align}
\label{eq:Gamma3}
    \Gamma^{ijkl}_3(\bm r) = \frac{\chi^{ijkl}_3(\bm r)}{\epsilon^2_0\epsilon^4(\bm r)},
\end{align}
where we have assumed that the relative dielectric function $\epsilon(\bm r)$ does not depend on frequency  over the range of frequencies within each ring resonance, and have neglected any dependence of $\chi_3^{ijlk}(\bm r)$ over the frequencies of interest.

Our process of interest is SFWM, where one
pump photon in resonance $P$ and one in resonance $C$ are annihilated, and two photons in
resonance $S$ are generated  (see Fig. \ref{fig:ring}(a)). The asymptotic-out fields for the actual and phantom waveguides are used for the generated photons, while only the asymptotic-in mode field for the actual waveguide is used for the pump photons; hence we can drop the index $n$ on the asymptotic-in mode fields and operators, understanding the actual input channel to be identified. If we consider only the terms responsible for SFWM in Eq. \eqref{eq:HNL}, the nonlinear Hamiltonian becomes \cite{MBanicRingLoss2022}
\begin{align}
\label{eq:HSFWM}
    H_{NL} &= -\frac{3}{\epsilon_0}\sum_{n,n'}\int dk_1dk_2dk_3dk_4 K^{nn'}(k_1,k_2,k_3,k_4) \nonumber
    \\
    &\times a^\dagger_{n S k_1}a^\dagger_{n' S k_2} b_{P k_3}b_{C k_4} + {\rm H.c.},
\end{align}
where the nonlinear coefficient for SFWM is given by
\begin{align}
    \label{eq:K}
    K^{nn'}(k_1,k_2,k_3,k_4) &= \int d\bm r\Gamma^{ijkl}_3(\bm r)  \left[ D^{\text{\textit{out}},i}_{n S k_1}(\bm r)\right]^*\nonumber
    \\
    &\times\left[D^{\text{\textit{out}},j}_{n' S k_2}(\bm r)\right]^* D^{\text{\textit{in}},k}_{P k_3}(\bm r)D^{\text{\textit{in}},l}_{C k_4}(\bm r).
\end{align}
We take the integral in Eq. \eqref{eq:K} to range over the ring where the fields are concentrated, and the effect of the nonlinearity is most significant.
The sums over $n$ and $n'$, which range over the actual
and phantom output channels, 
take into account the four combinations of two photons exiting the ring through two output channels. That is,  both photons can leave through the phantom or actual output channel, or one through each of the channels.
The nonlinear coefficient $K^{nn'}(k_1,k_2,k_3,k_4)$ gives the strength of the interaction corresponding to each combination. It is symmetric with respect to exchanging
the order of both $n$ and $n'$ and $k_1$ and $k_2$; that is,
\begin{align}
\label{eq:symmetric K}
     K^{nn'}(k_1,k_2,k_3,k_4) = K^{n'n}(k_2,k_1,k_3,k_4).
\end{align}
But it is not symmetric with respect to only exchanging the order of $n$ and $n'$ or $k_1$ and  $k_2$; that is,
\begin{align}
     K^{nn'}(k_1,k_2,k_3,k_4) \neq K^{n'n}(k_1,k_2,k_3,k_4).
\end{align}


\section{Input and output kets}
\label{sec:inoutstates}

The  unitary time-evolution operator for the system, $\mathcal{U}(t,t')$, is a solution to the Schr\"odinger equation
\begin{align}
\label{eq:dUdt initial}
    i\hbar\frac{d\mathcal{U}(t,t')}{dt} = H \mathcal{U}(t,t'),
\end{align}
satisfying $\mathcal{U}(t',t')=I$ for all $t'$, where $I$ is the identity operator. 
We write the  full Hamiltonian $H$ as the sum of the linear and nonlinear contributions,
\begin{align}
\label{eq:startH}
H = H_{L} + H_{NL},
\end{align}
where 
$H_{L}$  is given by Eq. \eqref{eq:HL} and $H_{NL}$ is given by Eq. \eqref{eq:HSFWM}. We assume that times $t_{on}$ and $t_{\textit{off}}$ can be identified such that:  (a) For $t<t_{on}$ the excitations of the quantum fields incident on the ring are sufficiently far from it that $H_{NL}$ has no effect on the evolution of the ket, and that (b) for $t>t_{\textit{off}}$ all the field excitation has propagated sufficiently far from the ring that, again, $H_{NL}$ has no effect on the evolution of the ket.

Consider then the evolution of a ket from some early time $t_0 < t_{\textit{on}}$ to some later time $t_1 >t_{\textit{off}}$. The evolution operator can be written as
\begin{align}
\label{eq:Ut0t1}
    \mathcal{U}(t_1,t_0) =  \mathcal{U}(t_1,t_{\textit{off}}) \mathcal{U}(t_{\textit{off}},t_{\textit{on}}) \mathcal{U}(t_{\textit{on}},t_0).
\end{align}
 The evolution operator from $t_0$ to $t_{\textit{on}}$ and  $t_{\textit{off}}$ to $t_1$ only involves the linear Hamiltonian, that is
\begin{align}
   \mathcal{U}(t_{\textit{on}},t_0) &= {\rm e}^{-iH_L(t_{\textit{on}} - t_0)/\hbar},
  \\
     \mathcal{U}(t_1,t_{\textit{off}}) &= {\rm e}^{-iH_L(  t_1-t_{\textit{off}})/\hbar}.
\end{align}
We can then write Eq. \eqref{eq:Ut0t1} as
\begin{align}
     \mathcal{U}(t_1,t_0) = {\rm e}^{-iH_L(  t_1-t_{\textit{off}})/\hbar} \mathcal{U}(t_{\textit{off}},t_{\textit{on}}){\rm e}^{-iH_L(t_{\textit{on}} - t_0)/\hbar}.
\end{align}

We now suppose that an initial ket $\ket{\psi(t_0)}$ is specified at a time $t_0<t_{\textit{on}}$. We define the ``input ket"
\begin{align}
\label{eq:inket}
\ket{\psi_{in}} \equiv {\rm e}^{-iH_L(0 - t_0)/\hbar} \ket{\psi(t_0)},
\end{align}
which is the ket that the initial ket would evolve to at $t=0$ were $H_{NL} = 0$. And starting with the ket $\ket{\psi(t_1)}$ characterizing the state at a time $t_1>t_{\textit{off}}$, we define  the ``output ket"
\begin{align}
\ket{\psi_{out}}\equiv{\rm e}^{iH_L(t_1- 0)/\hbar}\ket{\psi(t_1)},
\end{align}
which is the ket at $t=0$ that would evolve to  
$\ket{\psi(t_1)}$ at $t=t_1$ were $H_{NL} = 0$. Since the evolution of the ket from from $t_0$ to $t_1$ is given by
\begin{align}
\label{eq:psi01}
\ket{\psi(t_1)} =  \mathcal{U}(t_1,t_0)\ket{\psi(t_0)},
\end{align}
in terms of the input and output kets
this can be written as
\begin{align}
\ket{\psi_{out}} = U(t_1,t_0) \ket{\psi_{in}},
\end{align}
where we defined
\begin{align}
\label{eq:Utt}
U(t_1,t_0) \equiv {\rm e}^{iH_Lt_1/\hbar} \mathcal{U}(t_1,t_0) {\rm e}^{-iH_Lt_0/\hbar}.
\end{align}
Then taking the limit
\begin{align}
    U(\infty, -\infty) = \lim_{t_1 \rightarrow \infty, t_0 \rightarrow -\infty} U(t_1,t_0),
\end{align}
we obtain
\begin{align}
\ket{\psi_{out}} = U(\infty,-\infty) \ket{\psi_{in}}.
\end{align}
It is then convenient to define
\begin{align}
\label{eq:overline U}
    \overline{U}(t) \equiv U(t,-\infty),
\end{align}
for all times $t$. We can then write
\begin{align}
\ket{\psi_{out}} = |\overline{\psi}(\infty)\rangle,
\end{align}
where
\begin{align}
\label{eq:barpsi}
    |\overline{\psi}(t)\rangle &= \overline{U}(t) \ket{\psi_{in}},
\end{align}
and
\begin{align}
\label{eq:initial} 
\ket{\psi_{in}}=|\overline{\psi}(-\infty)\rangle. 
\end{align}
Taking the derivative of Eq. \eqref{eq:barpsi} and using Eqs. \eqref{eq:dUdt initial}, \eqref{eq:Utt}, and \eqref{eq:overline U} we obtain
\begin{align}
\label{eq:psibardot}
i \hbar \frac{d}{dt}  |\overline{\psi}(t)\rangle = \overline{H}_{NL}(t) |\overline{\psi}(t)\rangle, 
\end{align}
where we have defined the interaction picture nonlinear Hamiltonian as
\begin{align}
\label{eq:interaction H}
\overline{H}_{NL}(t)\equiv  {\rm e}^{iH_Lt/\hbar} H_{NL}  {\rm e}^{-iH_Lt/\hbar},
\end{align}
and as $t$ ranges from $-\infty$ to $\infty$ in Eq. \eqref{eq:psibardot} the ket evolves from $\ket{\psi_{in}}$ to $\ket{\psi_{out}}$.

\section{Unitary evolution}
\label{sec:Ubar}
In Appendix A we simplify the nonlinear Hamiltonian Eq. \eqref{eq:interaction H} by treating the strong CW pump classically, and discretize the $k$ involved for the weak pump and generated fields to implement numerical calculations. The result can be written as
\begin{align}
\label{eq:Hint spdc discrete 3}
    \overline{H}_{NL}(t)&=  \hbar \sum_{\mu, \nu}\sum_{l}\Lambda_{\mu \nu l}(t) a^\dagger_{\mu}a^\dagger_{\nu} b_{l} + {\rm H.c.},
\end{align}
where for convenience we removed the discrete labels $P$ and $S$ on the input and output operators, and they  satisfy the usual commutation relations $[b_{l}, b^\dagger_{l'}] = \delta_{ll'}$  and $[a_{\mu}, a^\dagger_{\nu}] = \delta_{\mu\nu}$. Here $\Lambda_{\mu \nu l}(t)$ is the nonlinear coefficient for the effective $\chi^{(2)}$ interaction, where $l$ labels a discrete wavenumber $k_l$ for the pump, and $\mu$ and $\nu$ are discrete indices for the generated photons (see Eq. \eqref{eq:lambda ring}). Each Greek index $\mu$, for example, refers to two properties of the photon, its discrete wavenumber $k_i$ and the output channel $n$. The sum over $\mu$ in Eq. \eqref{eq:Hint spdc discrete 3} represents two sums; one over the actual and phantom output channel and the second over the discrete wavenumbers $k_i$. 

We now turn to the unitary evolution in Eq. \eqref{eq:psibardot}. Here we do not consider a general ket $\ket{\psi_{in}}$, but restrict ourselves to
an input ket that is a coherent state in the actual input channel and vacuum in all other channels,
\begin{align}
\label{eq:inket coherent}
    \ket{\psi_{in}} &= D_b(\bm \beta_{in})\ket{vac},
\end{align}
where  $D_b(\bm \beta(t))$ is the unitary displacement operator, with displacement parameter $\bm \beta_{in} \equiv \bm \beta(-\infty)$. It is written in terms of the input operators as
\begin{align}
\label{eq:Db}
    D_b(\bm \beta(t)) = \exp([\bm \beta(t)]^{\rm T}\bm b^\dagger - [\bm \beta^*(t)]^{\rm T}\bm b),
\end{align}
where $\bm \beta(t) = [\beta_1(t), \beta_2(t),\ldots]^{\rm T}$ and $\bm \beta^*(t) = [\beta^*_1(t), \beta^*_2(t),\ldots]^{\rm T}$  are column vectors of displacement parameters, and $\bm b = [b_1, b_2, \ldots]^{\rm T}$ and $\bm b^\dagger = [b^\dagger_1, b^\dagger_2, \ldots]^{\rm T}$ are column vectors of input operators.  We then seek a solution of \eqref{eq:psibardot}
of the form 

\begin{align}
\label{eq:psibarform}   |\overline{\psi}(t)\rangle=U_a(t)U_b(t)|\tilde{\psi}(t)\rangle, 
\end{align}
where $|\tilde{\psi}(t)\rangle$ describes the effect of a non-Gaussian unitary operator acting on the vacuum state, as discussed below, and
$U_a(t)$ and $U_b(t)$ are Gaussian unitary operators, given by
\begin{align}
\label{eq:Ua}
    U_a(t) &= S_a(\bm J(t))R_a(\bm\phi(t)){\rm e}^{i\theta_a(t)},
    \\
    \label{eq:Ub}
    U_b(t) &= D_b(\bm \beta(t)){\rm e}^{i\theta_b(t)},
\end{align}
where $\theta_b(t)$ and $\theta_a(t)$ are real functions of time,  $S_a(\bm J(t))$ is a multimode squeezing operator, and   $R_a(\bm \phi(t))$ is a multimode rotation operator.

The unitary squeezing operator $S_a(\bm J(t))$ is given by \cite{marhodes1990}
 \begin{align}
\label{eq:S}
     S_a(\bm J(t)) &= \exp(\frac{1}{2} {\bm a^\dagger}^{\rm T}\bm J(t) {\bm a^\dagger} - {\rm H.c.}),
 \end{align}
where $\bm a = [a_1, a_2, \ldots]^{\rm T}$ and $\bm a^\dagger = [a^\dagger_1, a^\dagger_2, \ldots]^{\rm T}$ are column vectors of output operators and $\bm J(t)$ is the symmetric time-dependent squeezing matrix, with $\bm J(t) = \bm J^{\rm T}(t)$. The unitary rotation operator $R_a(\bm \phi(t))$ is given by \cite{marhodes1990}
 \begin{align}
     \label{eq:R}
     R_a(\bm \phi(t)) &= \exp(i {\bm a^\dagger}^{\rm T}\bm \phi(t) \bm a),
\end{align}
where $\bm \phi(t)$ is
the Hermitian time-dependent rotation matrix,
with $\bm \phi (t) = \bm \phi^\dagger(t)$. The parameters $\bm \beta(t)$, $\bm J(t)$, and $\bm \phi(t)$ are subject to the following initial conditions $\bm \beta(-\infty) = \bm \beta_{in}$, $\bm J(-\infty) = 0$, and $\bm \phi(-\infty) = 0$, and also $\theta_b(-\infty) = \theta_a(-\infty) = 0$; further, we 
take
\begin{align}
\label{eq:starttilde}
   |\tilde{\psi}(-\infty)\rangle = |vac\rangle. 
\end{align}
This guarantees that the initial condition \eqref{eq:initial} is satisfied.

Returning to Eq. \eqref{eq:psibarform}, our goal is to define the parameters in the Gaussian unitary operators in Eq. \eqref{eq:Ua} and Eq. \eqref{eq:Ub} so that the ket $|\tilde{\psi}(t)\rangle$ isolates the non-Gaussian behavior of the full ket in Eq. \eqref{eq:psibarform}. Now regardless of how the parameters in those Gaussian unitary operators are chosen to depend on time, we find that for Eq. \eqref{eq:psibarform} to satisfy the full Schr\"odinger equation in Eq. \eqref{eq:psibardot} we require
\begin{align}
\label{eq:tildedynamics}
    i \hbar \frac{d}{dt} |\tilde{\psi}(t)\rangle = H_{\textit{eff}}(t) |\tilde{\psi}(t)\rangle, 
\end{align}
where we have introduced the effective Hamiltonian
\begin{align}
\label{eq:Heff}
H_{\textit{eff}}(t)&=H_I(t) + U^\dagger_a(t)H_a(t)U_a(t) -  i\hbar U^\dagger_a(t)\frac{dU_a(t)}{dt}\nonumber
\\
&+U^\dagger_b(t)H_b(t) U_b(t) -i\hbar U^\dagger_b(t)\frac{dU_b(t)}{dt},
\end{align}
with
\begin{align}
\label{eq:Ha}
    H_a(t) &\equiv \hbar\Lambda_{\mu \nu l}(t)[\bm \beta(t)]_la^\dagger_\mu a^\dagger_\nu + {\rm H.c.},
    \\
    \label{eq:Hb}
    H_b(t)&\equiv \hbar  \Lambda^*_{\mu \nu l}(t)  [\bm V(t)\bm W^{\rm T}(t)]_{\mu \nu} \left(b^\dagger_l - [\bm \beta^*(t)]_l\right)+ {\rm H.c.},
    \\
    \label{eq:HI}
H_I(t) &\equiv\hbar  \Lambda^*_{\mu \nu l}(t) \bigg( [\bm V(t)]_{\mu \mu'} [\bm V(t)]_{\nu \nu'} a_{\mu'} a_{\nu'} \nonumber
\\
&+ \left([\bm V(t)]_{\mu \nu'} [\bm W(t)]_{\nu \mu'}  + [\bm W(t)]_{\mu \mu'} [\bm V(t)]_{\nu \nu'}\right) a^\dagger_{\mu'}a_{\nu'} \nonumber
\\
&+ [\bm W(t)]_{\mu \mu'} [\bm W(t)]_{\nu \nu'} a^\dagger_{\mu'}a^\dagger_{\nu'} \bigg)b^\dagger_l  +{\rm H.c.}.
\end{align}
Here repeated indices are summed over, and the matrices $\bm{V}(t)$ and $\bm{W}(t)$ are defined in terms of the parameters appearing in the Gaussian unitary operators in Eq. \eqref{eq:Ua} and Eq. \eqref{eq:Ub}; see Appendix B for details.

The Hamiltonians appearing above each capture part of the physics: 
$H_a(t)$ describes the generation of photon pairs by an effective $\chi^{(2)}$ interaction modulated by a time-dependent pump amplitude $\bm \beta(t)$; 
$H_b(t)$ determines the rate of depletion of the pump amplitude $\bm \beta(t)$, which is being driven by the generation of photon pairs; 
$H_I(t)$, which  only contains non-Gaussian terms, describes the quantum correlations between the pump and the generated light. 

To achieve our goal of moving all the non-Gaussian behavior to $|\tilde{\psi}(t)\rangle$, we must choose the time dependence of the parameters appearing in the Gaussian operators in Eq. \eqref{eq:Ua} and Eq. \eqref{eq:Ub} so that
 \begin{align}
 \label{eq:Hefffinal}
H_{\textit{eff}}(t) =H_I(t).
\end{align}
It is easy to show that this condition is satisfied if we have
 \begin{align}
\label{eq:Uadot}
 i\hbar \frac{dU_a(t)}{dt} &=H_{a}(t)U_a(t),
 \\
 \label{eq:Ubdot}
i\hbar \frac{dU_b(t)}{dt}  &= H_{b}(t)U_b(t).
 \end{align}   
We show in Appendix \ref{sec:Heff simp} that this can be achieved by
appropriately choosing the parameters $\bm V(t)$, $\bm W(t)$, $\bm \beta(t)$, and $\theta_{b}(t)$ so that
\begin{align}
\label{eq:Vdot}
    \frac{d\bm V(t)}{dt}&=-2i\bm \zeta(t)\bm W^*(t),
    \\
    \label{eq:Wdot}
       \frac{d\bm W(t)}{dt}&=-2i\bm \zeta(t)\bm V^*(t),
       \\
       \label{eq:dot beta}
    i\frac{d[{\bm \beta}(t)]_l}{dt}  &= [\bm \gamma(t)]_l, 
    \\
    \label{eq:dot thetab}
    \theta_b(t)&=\frac{1}{2}\int_{-\infty}^tdt'[\bm \gamma(t')]_l[{\bm \beta}^*(t')]_l + {\rm c.c.},
\end{align}
are satisfied, where the matrix $\bm \zeta(t)$ drives the production of photon pairs in the squeezed state,
\begin{align}
    \label{eq:zeta}
[\bm \zeta(t)]_{\mu \nu} =  \Lambda_{\mu \nu l}(t)[\bm \beta(t)]_l,
\end{align}
and the vector $\bm \gamma(t)$ determines the rate of pump depletion,
\begin{align}
\label{eq:gamma(t)}
    [\bm \gamma(t)]_l = \Lambda^*_{\mu \nu l}(t)  [\bm V(t)\bm W^{\rm T}(t)]_{\mu \nu}.
\end{align}
 
 We solve the Eqs. \eqref{eq:Vdot}, \eqref{eq:Wdot}, and \eqref{eq:dot beta} numerically for the parameters $\bm V(t)$, $\bm W(t)$, and $\bm \beta(t)$, which are subject to the initial conditions $\bm \beta(-\infty) = \bm \beta_{in}$,  $\bm W(-\infty) = 0$, and $\bm V(-\infty) = \bm I$, where $\bm I$ is the identity matrix. The solutions are 
 put into Eq. \eqref{eq:dot thetab}, which is then integrated to obtain the phase $\theta_b(t)$. In Appendix \ref{sec:extract J} we show how to extract the squeezing matrix $\bm J(t)$ and rotation matrix $\bm \phi(t)$ from the solutions $\bm V(t)$ and $\bm W(t)$, and in Appendix \ref{sec:theta} we give an equation for the phase $\theta_a(t)$ in Eq. \eqref{eq:Ua}. This procedure provides a solution for the Gaussian parameters of the unitary operators in Eqs. \eqref{eq:Ua} and \eqref{eq:Ub}, and is valid for high pump power and nonlinear interaction strength. 

Still necessary for a full solution, of course, is to include the non-Gaussian evolution of $|\tilde{\psi}(t)\rangle$. Then the full solution can be written as the squeezed, displaced, and rotated version of non-Gaussian terms acting only on the vacuum state,
\begin{align}
\label{eq:psibarformn}   |\overline{\psi}(t)\rangle=U_a(t)U_b(t)|\tilde{\psi}(t)\rangle, 
\end{align}    
where recall Eq. \eqref{eq:starttilde}. An attractive feature of this approach is that we can work in the Gaussian limit by setting 
$|\tilde{\psi}(t)\rangle \rightarrow |vac\rangle$, and still study the effects of pump depletion and the dynamics of the joint spectral amplitude in squeezed state generation.

That is what we do for 
the remainder of this
paper. We put 
\begin{align}
    |\overline{\psi}(t)\rangle \rightarrow U_a(t) U_b(t) |vac\rangle \equiv |\overline{\psi}_G(t)\rangle,
    \label{eq:Gdef}
\end{align}
and we have
\begin{align}
\label{eq:Gaussian ket}
 \ket{\overline{\psi}_G(t)} =  {\rm e}^{i\theta_a(t)}S_a(\bm J(t))\ket{vac} \otimes  {\rm e}^{i\theta_b(t)}\ket{\bm \beta(t)}.
\end{align}
where  $\ket{\bm \beta(t)}=D_b(\bm \beta(t)) \ket{vac}$ is a coherent state.
The Gaussian state is a product of a squeezed vacuum state written in terms of the output operators and a coherent state written in terms of the input operators, and from \eqref{eq:Gdef} its dynamics is governed by 
\begin{align}
\label{eq:appevolution}
     i\hbar \frac{d \ket{\overline{\psi}_G(t)}}{dt} = \left(H_a(t)+H_b(t)\right) \ket{\overline{\psi}_G(t)}.
\end{align}
This of course is an approximate result, and there are consequences. For example,
the operator $Q = {\bm a^\dagger}^{\rm T} \bm a /2+ {\bm b^\dagger}^{\rm T} \bm b$ commutes with $H_L$ and $H_{NL}$, and therefore with the full Hamiltonian in Eq. \eqref{eq:startH}, and so it is a conserved quantity. 
It therefore also commutes with $\overline{H}_{NL}(t)$, which governs the evolution of the full $|\overline{\psi}(t)\rangle$ according to Eq. \eqref{eq:psibardot}. 
However, it does \textit{not} commute with the Hamiltonian $H_a(t)+H_b(t)$ that governs the evolution of the its approximation $|\overline{\psi}_G(t)\rangle.$ Nonetheless, in Appendix \ref{sec:conservedQ} we show that $\langle \overline{\psi}_G(t)|Q|\overline{\psi}_G(t) \rangle$ \textit{is} constant as $\ket{\overline \psi_G(t)}$ evolves according to Eq. \eqref{eq:appevolution}.


\section{First-order solution}
\label{sec:low}
If the energy of the pump is sufficiently low or the nonlinear coefficients $\Lambda_{\mu \nu l}(t)$ are sufficiently weak, then the elements of the matrix $\bm \zeta(t)$ in Eq. \eqref{eq:zeta} will be much less than unity, 
$[\bm \zeta(t)]_{\mu\nu}\ll 1$ for all $\mu$ and $\nu$. In this case the solution to Eq. \eqref{eq:Vdot} and Eq. \eqref{eq:Wdot} can be approximately written as \cite{beyondPhotonPairs}
\begin{align}
\label{eq:Vapprox}
    \bm V(t) &\approx \bm I,
    \\
    \label{eq:Wapprox}
        \bm W(t) &\approx  -2i\int_{-\infty}^t dt' \bm \zeta(t').
\end{align}
Consequently,  the pump depletion rate in Eq. \eqref{eq:dot beta} is negligible
\begin{align}
\label{eq:beta dot approx}
   \frac{d[{\bm \beta}(t)]_l}{dt} &\approx -2i \Lambda^*_{\mu \nu l}(t)\int_{-\infty}^t dt'[\bm \zeta(t')]_{\mu\nu}\approx 0,
\end{align}
since it depends on the product of the 
nonlinear coefficients, which are assumed to be small. The Hamiltonian  $H_b(t)$ in Eq. \eqref{eq:Hb}  can be set to zero, 
$H_b(t) \rightarrow 0$, since the coefficients in $H_b(t)$ are given by $d[{\bm \beta}(t)]_l/dt$. Similarly the phase $\theta_b(t)$ in  Eq. \eqref{eq:dot thetab}  is negligible, $\theta_b(t) \approx 0$.

With these approximations, the Gaussian ket $|\overline{\psi}_G(t)\rangle$ in Eq. \eqref{eq:Gaussian ket} can be written as
\begin{align}
    \label{eq:Gaussian ket low gain}
 \ket{\overline{\psi}_G(t)} = S_a(\bm J(t))\ket{vac} \otimes \ket{\bm \beta_{in}},
\end{align}
which satisfies the Schr\"odinger equation
\begin{align}
\label{eq:SE low gain}
    i\hbar \frac{d \ket{\overline{\psi}_G(t)}}{dt} = H_a(t) \ket{\overline{\psi}_G(t)}.
\end{align}
Treating $H_a(t)$ as a small quantity we can construct a first-order solution to  Eq. \eqref{eq:SE low gain} using the Magnus expansion \cite{Blanes2009}, and keeping only the first term we obtain
\begin{align}
\label{eq:Gaussian ket magnus}
    \ket{\overline{\psi}_G(t)} \approx \exp(-i\int_{-\infty}^tdt' [\bm \zeta(t')]_{\mu\nu} a^\dagger_\mu a^\dagger_\nu  - {\rm H.c.}) \otimes \ket{\bm \beta_{in}}.
\end{align}
The higher-order terms in the expansion involve the commutator between the Hamiltonian at two different times, such as $[H_a(t), H_a(t')]$ \cite{Blanes2009}. The ket in Eq. \eqref{eq:Gaussian ket magnus} is the solution when we assume that $[H_a(t), H_a(t')] = 0$ so that time-ordering effects are negligible. It was pointed out earlier for waveguides \cite{Yanagimoto2022, beyondPhotonPairs} that this assumption becomes invalid when the power of the pump is increased. We  demonstrate 
below that this also holds for a squeezed state generated in a ring resonator. Comparing Eq. \eqref{eq:Gaussian ket magnus} and Eq. \eqref{eq:Gaussian ket low gain} we immediately obtain the approximate first-order solution for the squeezing matrix
\begin{align}
    \label{eq:J lowgain}
    \bm J(t) \approx -2i\int_{-\infty}^t dt' \bm \zeta(t').
\end{align}
This is the expression for $\bm W(t)$ in Eq. \eqref{eq:Wapprox}, so we can write $\bm J(t) \approx \bm W(t)$. To obtain the form in Eq. \eqref{eq:Gaussian ket low gain} we have put $\theta_a(t) = 0$, which is valid for a weak nonlinearity (see Appendix \ref{sec:theta}).


\section{Example: Squeezed state generation in a ring resonator}
\label{sec:numericalring}
In this section we study the example of generating a highly squeezed state in a ring resonator. We calculate the squeezing matrix, number of generated photons, Schmidt number, and the second-order correlation function in the actual output channel including scattering loss, and compare these results to the first-order solution.

We consider a SiN ring resonator point-coupled to an actual waveguide and a
phantom waveguide, the latter modelling scattering loss; see
Fig. \ref{fig:ring}(b).  For a sample calculation we consider a structure with
a ring radius of $R = 64 \mu$m, $\omega_S = 2\pi \times 193 {\rm THz}$, a group velocity of $1.5\times 10^8 {\rm m/s}$, a nonlinear parameter of $\overline{\gamma}_{nl}\approx 1({\rm Wm})^{-1}$ (see Eq. \eqref{eq:gammanl SFWM}), and an intrinsic quality factor of $2\times 10^6$ for all three ring resonances. We take the loaded quality factors for the $P$, $S$, and $C$ modes to be $4\times10^4$\,(over-coupled), $2\times 10^6$\,(over-coupled), and $1\times 10^6$\,(critically-coupled), respectively. 

In mode $C$ there is an undepleted CW pump with power $P_C = 30{\rm mW}$, and in mode $P$ there is a weak pulsed pump with energy $U_P$ and duration $\tau_P = 0.5{\rm ns}$. The initial amplitude of the pulse is
\begin{align}
\label{eq:gaussian betain}
    [\bm \beta_{in}]_l = \sqrt{\frac{N_P v_P\tau_P}{\sqrt{\pi}}}\exp(-\frac{1}{2}(k_l-K_P)^2v_P^2\tau_P^2),
\end{align}
where for simplicity we neglect group velocity dispersion, $N_P$ is the initial photon number, $v_p$ is the group velocity, and $K_P$ is the central wavenumber.

\begin{figure}[htbp]
    \centering
    \includegraphics[scale = 0.4]{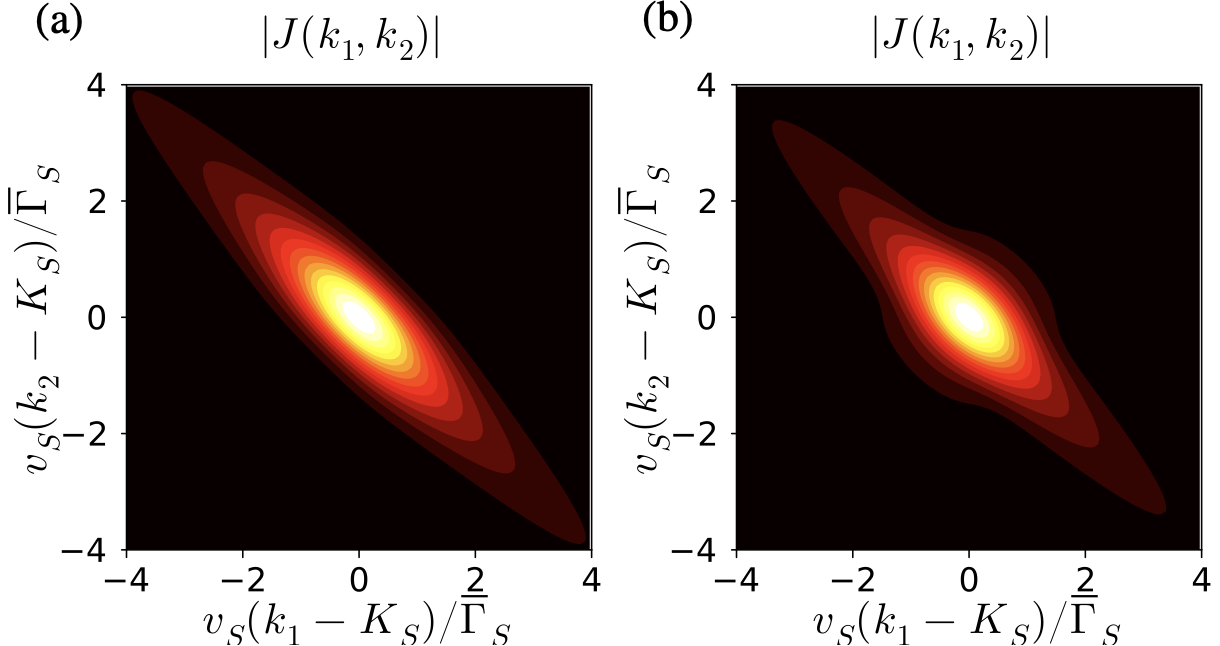}
    \caption{ Magnitude of the squeezing matrix $|J(k_1,k_2)|$ in the high-gain regime ($U_P = 100{\rm pJ}$) calculated using the  (a) first-order solution and the (b) full Gaussian solution.}
    \label{fig:J}
\end{figure}

In Fig. \ref{fig:J}(a) the magnitude of the squeezing matrix is shown for $U_P = 100{\rm pJ}\,(N_P\approx7.8\times 10^8)$. The x-axis and y-axis represent the detuning from the center of the ring resonance of the mode $S$, where $v_S$, $K_S$, and $1/\overline{\Gamma}_S\approx 330\,{\rm ps}$  are its group velocity, central wavenumber, and dwelling time, respectively. For this energy the average generated photon number in the squeezed state is about 19, which corresponds to approximately $19\,{\rm dB}$ of squeezing, putting it in the high-gain regime.  

The squeezing matrix in Fig. \ref{fig:J}(a) is calculated using the approximate first-order solution in Eq. \eqref{eq:J lowgain}. In order to compare the first-order solution to the full Gaussian solution, in Fig. \ref{fig:J}(b) the magnitude of the squeezing matrix is shown for the same parameters as above but 
calculated using the full Gaussian solution in Eq. \eqref{eq:J matrix} that is valid for arbitrary energy. The full Gaussian solution in Fig. \ref{fig:J}(b) exhibits less correlation between the wavenumbers ($k_1$ and $k_2$) of the generated photon pairs than the first-order solution in Fig. \ref{fig:J}(a).

\begin{figure}[htbp]
    \centering
    \includegraphics[scale=0.4]{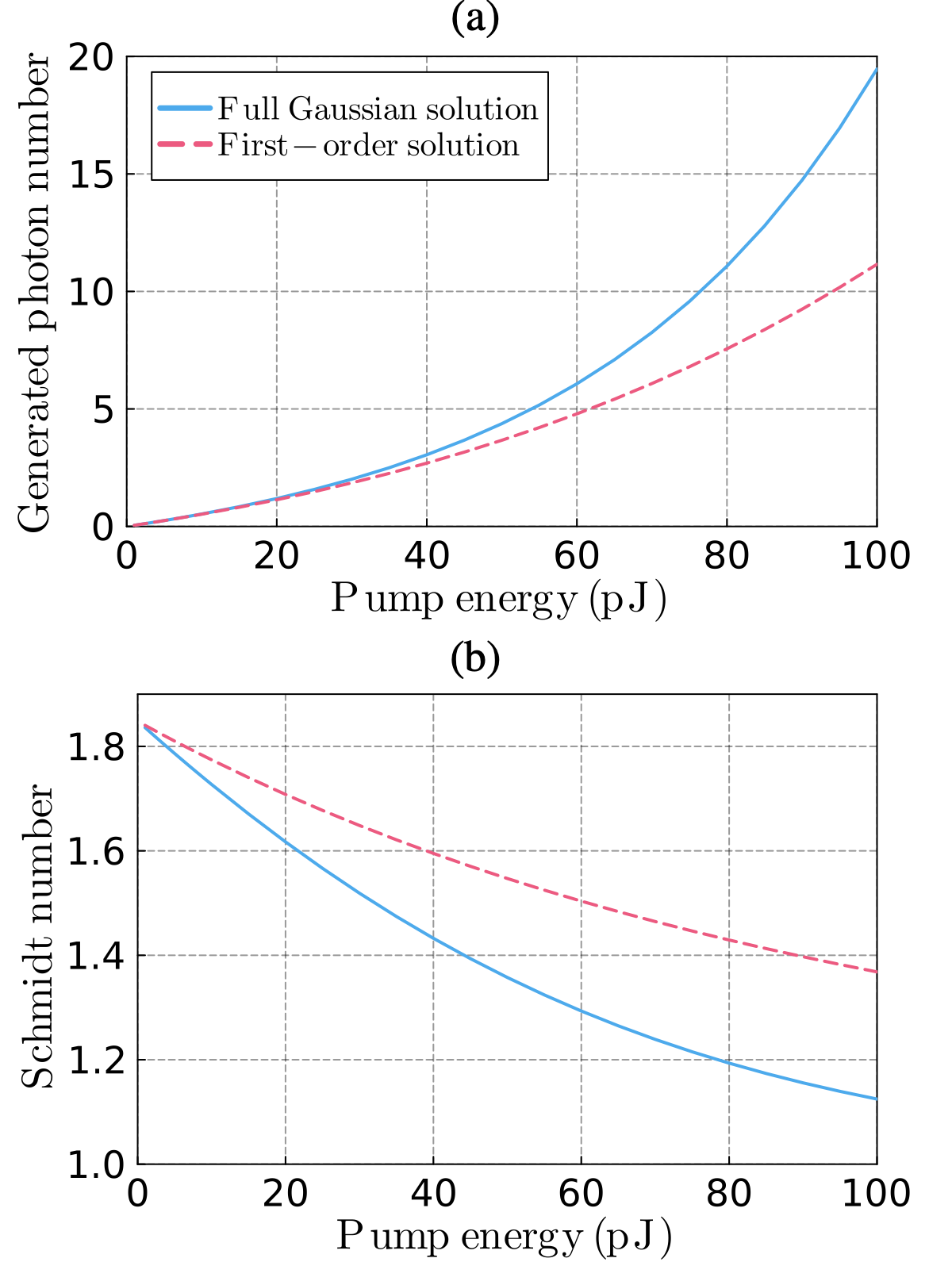}
    \caption{(a) The average generated photon number in the squeezed state and (b) the Schmidt number, calculated with the full Gaussian solution (blue line) and the first-order solution (pink dashed) for increasing energy of the pulsed pump ($U_P$).}
    \label{fig:Schmidtnumber}
\end{figure}

In Fig. \ref{fig:Schmidtnumber}(a) the average generated photon number in the squeezed state is shown for pump energy ($U_P$) between $1\,{\rm pJ}$ and $100\,{\rm pJ}$. The calculation is done with the full Gaussian solution (blue line) and the first-order solution (pink dashed). The photon number is calculated with ${\rm tr}(\bm W^*(t) \bm W^{\rm T}(t))$, where the matrix $\bm W(t)$ is obtained using the full Gaussian solution or the first-order solution. The first-order solution gives a photon number that is approximately $20\%$ less than the full Gaussian solution for pump energies at or below $60\,{\rm pJ}$. This corresponds to an underestimation of the squeezing by about $1\,{\rm dB}$ at most. But at $80\,{\rm pJ}$ and $100\,{\rm pJ}$ the first-order solution underestimates the  squeezing by $1.6\,{\rm dB}$ and $2.3\,{\rm dB}$, respectively. Despite the high level of squeezing, the level of pump depletion is negligible. As a result the neglect of the non-Gaussian corrections to the vacuum, which are captured by the evolution Eq. \eqref{eq:tildedynamics} of $|\tilde{\psi}(t) \rangle$ -- and which we expect to only become important when there is significant pump depletion \cite{Yanagimoto2022} -- is justified. 

In Fig. \ref{fig:Schmidtnumber}(b) the Schmidt number \cite{Houde2023} of the squeezing matrix is shown for the same parameters as above. The first-order solution consistently overestimates the Schmidt number, and thus
the squeezing matrix is more separable than what is predicted by the first-order solution.

\begin{figure}
    \centering
    \includegraphics[scale = 0.4]{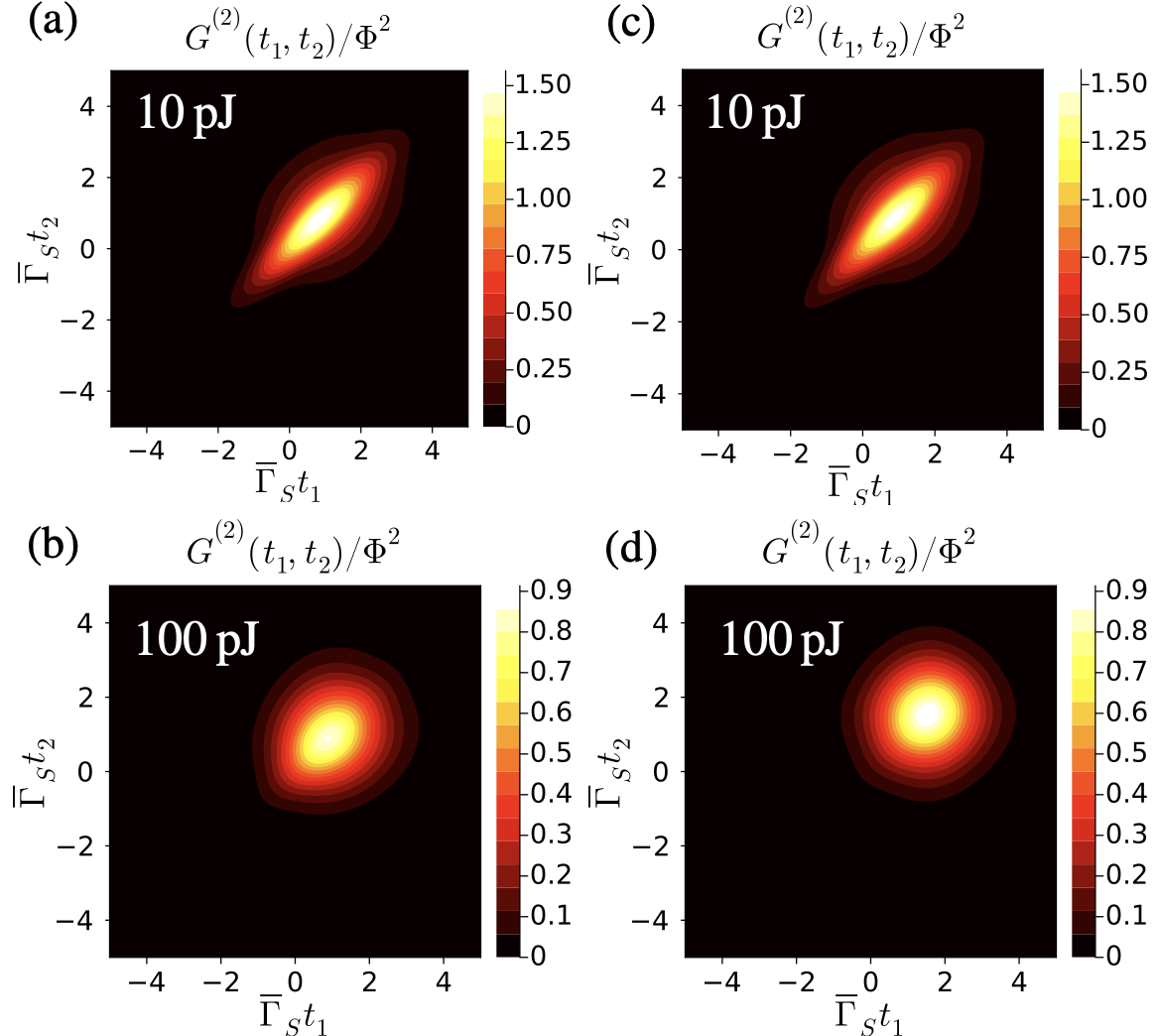}
    \caption{Comparison of the second-order correlation functions $G^{(2)}(t_1,t_2)$ in the actual output channel using the first-order solution (left column) with (a) $U_P = 10\,{\rm pJ}$ and (b) $U_P = 100\,{\rm pJ}$, as well as using the full Gaussian solution (right column) with (c) $U_P = 10\,{\rm pJ}$ and (d) $U_P = 100\,{\rm pJ}$.}
    \label{fig:G2}
\end{figure}

Each Schmidt mode typically involves light exiting into both the actual and the phantom waveguide, the latter modelling the scattering of light off the chip. To focus on the properties of the light exiting the actual waveguide, we
consider a measurement  of the second-order correlation function $G^{(2)}(t_1,t_2)$ of the squeezed light exiting the actual output channel. It is convenient to introduce a channel operator $\bar{a}(x,t)$ as the Fourier transform of the output operators for the actual waveguide $a_{k}$ as \cite{seifoory2022}
\begin{align}
    \label{eq:real out ch op}
    \bar{a}(x,t) = \int \frac{dk}{\sqrt{2\pi}} a_{k}{\rm e}^{i(k-K_S)(x-v_St)},
\end{align}
where $x$ is the direction of propagation in the waveguide. The operator $\bar{a}(x,t)$ is associated with an electric field at the position $x$ and time $t$.  Just beyond the coupling point between the ring and waveguide ($x=0^+$), the second-order correlation function is given by \cite{drago2023}
\begin{align}
\label{eq:G2}
    G^{(2)}(t_1,t_2) = v_S\langle \bar{a}^\dagger(0^+,t_1)\bar{a}^\dagger(0^+,t_2) \bar{a}(0^+,t_2)\bar{a}(0^+,t_1)\rangle,
\end{align}
where the expectation value is done with the Gaussian ket in Eq. \eqref{eq:Gaussian ket}. This can be used to predict the probability of detecting coincidence counts at the two times $t_1$ and $t_2$. We normalize it using $\Phi^2$, which is defined as the ratio of the number of generated photons per pump pulse duration, given by 
\begin{align}
    \Phi \equiv \langle n_S\rangle/\tau_P,
\end{align}
where $\langle n_S\rangle$ is the number of generated photons.  

In Figs. \ref{fig:G2}(a) and \ref{fig:G2}(b) the second-order correlation function for the actual output channel is shown using the first-order solution, as well as in Figs. \ref{fig:G2}(c) and \ref{fig:G2}(d) using the full Gaussian solution. When the energy of the pump is $10\,{\rm pJ}$ the correlation functions from the first-order solution and full Gaussian solution appear to agree. For the high energy pump at $100\,{\rm pJ}$ the first-order solution begins to show correlations that are not present in the full Gaussian solution, and the maximum of the first-order solution occurs at an early time ($t_1 = t_2 = 0.9/\overline{\Gamma}_S$) than the full Gaussian solution ($t_1 = t_1 = 1.5/\overline{\Gamma}_S$).

\section{Conclusion}
\label{sec:conclude}
In conclusion, we presented a multimode theory of Gaussian state generation for an effective $\chi^{(2)}$ interaction, in terms of asymptotic-in and -out fields, that includes scattering loss and pump depletion. We showed that the full ket for our system can be written as a Gaussian unitary acting on non-Gaussian
ket that satisfies a Schr\"odinger equation with an effective Hamiltonian that describes the quantum correlations between the pump and the generated light. In order for this to be valid, we required that the squeezing, rotation, and displacement parameters of the Gaussian unitary were solutions to a set of coupled differential equations.  As an example we studied the lossy generation of a highly squeezed state using an effective $\chi^{(2)}$ interaction in a silicon nitride ring resonator, and we compared this full Gaussian solution to a lowest-order Magnus expansion for the ket. In future work, we will study the evolution of the  non-Gaussian  ket, the effects of self- and cross-phase modulation, and the application to more general resonant systems. This work opens up a path to study the deterministic generation of non-Gaussian states in resonant systems.

\section*{Acknowledgements}
We thank Marco Liscidini for valuable discussions. J. E. S. and C. V. acknowledge the Horizon-Europe research and innovation program under grant agreement ID: 101070168 (HYPERSPACE) and Natural  Sciences  and  Engineering  Research Council of Canada (NSERC) for funding.

\bibliography{References-minimum}

\begin{thebibliography}{26}%
\makeatletter
\providecommand \@ifxundefined [1]{%
 \@ifx{#1\undefined}
}%
\providecommand \@ifnum [1]{%
 \ifnum #1\expandafter \@firstoftwo
 \else \expandafter \@secondoftwo
 \fi
}%
\providecommand \@ifx [1]{%
 \ifx #1\expandafter \@firstoftwo
 \else \expandafter \@secondoftwo
 \fi
}%
\providecommand \natexlab [1]{#1}%
\providecommand \enquote  [1]{``#1''}%
\providecommand \bibnamefont  [1]{#1}%
\providecommand \bibfnamefont [1]{#1}%
\providecommand \citenamefont [1]{#1}%
\providecommand \href@noop [0]{\@secondoftwo}%
\providecommand \href [0]{\begingroup \@sanitize@url \@href}%
\providecommand \@href[1]{\@@startlink{#1}\@@href}%
\providecommand \@@href[1]{\endgroup#1\@@endlink}%
\providecommand \@sanitize@url [0]{\catcode `\\12\catcode `\$12\catcode `\&12\catcode `\#12\catcode `\^12\catcode `\_12\catcode `\%12\relax}%
\providecommand \@@startlink[1]{}%
\providecommand \@@endlink[0]{}%
\providecommand \url  [0]{\begingroup\@sanitize@url \@url }%
\providecommand \@url [1]{\endgroup\@href {#1}{\urlprefix }}%
\providecommand \urlprefix  [0]{URL }%
\providecommand \Eprint [0]{\href }%
\providecommand \doibase [0]{https://doi.org/}%
\providecommand \selectlanguage [0]{\@gobble}%
\providecommand \bibinfo  [0]{\@secondoftwo}%
\providecommand \bibfield  [0]{\@secondoftwo}%
\providecommand \translation [1]{[#1]}%
\providecommand \BibitemOpen [0]{}%
\providecommand \bibitemStop [0]{}%
\providecommand \bibitemNoStop [0]{.\EOS\space}%
\providecommand \EOS [0]{\spacefactor3000\relax}%
\providecommand \BibitemShut  [1]{\csname bibitem#1\endcsname}%
\let\auto@bib@innerbib\@empty
\bibitem [{\citenamefont {Schnabel}(2016)}]{schnabelSqueezeStates}%
  \BibitemOpen
  \bibfield  {author} {\bibinfo {author} {\bibfnamefont {R.}~\bibnamefont {Schnabel}},\ }\bibfield  {title} {\bibinfo {title} {Squeezed states of light and their applications in laser interferometers},\ }\href@noop {} {\bibfield  {journal} {\bibinfo  {journal} {Phys. Rep.}\ }\textbf {\bibinfo {volume} {684}},\ \bibinfo {pages} {1} (\bibinfo {year} {2016})}\BibitemShut {NoStop}%
\bibitem [{\citenamefont {Weedbrook}\ \emph {et~al.}(2012)\citenamefont {Weedbrook}, \citenamefont {Pirandola}, \citenamefont {Garc\'{\i}a-Patr\'on}, \citenamefont {Cerf}, \citenamefont {Ralph}, \citenamefont {Shapiro},\ and\ \citenamefont {Lloyd}}]{Weedbrook2012}%
  \BibitemOpen
  \bibfield  {author} {\bibinfo {author} {\bibfnamefont {C.}~\bibnamefont {Weedbrook}}, \bibinfo {author} {\bibfnamefont {S.}~\bibnamefont {Pirandola}}, \bibinfo {author} {\bibfnamefont {R.}~\bibnamefont {Garc\'{\i}a-Patr\'on}}, \bibinfo {author} {\bibfnamefont {N.~J.}\ \bibnamefont {Cerf}}, \bibinfo {author} {\bibfnamefont {T.~C.}\ \bibnamefont {Ralph}}, \bibinfo {author} {\bibfnamefont {J.~H.}\ \bibnamefont {Shapiro}},\ and\ \bibinfo {author} {\bibfnamefont {S.}~\bibnamefont {Lloyd}},\ }\bibfield  {title} {\bibinfo {title} {Gaussian quantum information},\ }\href {https://doi.org/10.1103/RevModPhys.84.621} {\bibfield  {journal} {\bibinfo  {journal} {Rev. Mod. Phys.}\ }\textbf {\bibinfo {volume} {84}},\ \bibinfo {pages} {621} (\bibinfo {year} {2012})}\BibitemShut {NoStop}%
\bibitem [{\citenamefont {Wenger}\ \emph {et~al.}(2004)\citenamefont {Wenger}, \citenamefont {Tualle-Brouri},\ and\ \citenamefont {Grangier}}]{Wenger2004}%
  \BibitemOpen
  \bibfield  {author} {\bibinfo {author} {\bibfnamefont {J.}~\bibnamefont {Wenger}}, \bibinfo {author} {\bibfnamefont {R.}~\bibnamefont {Tualle-Brouri}},\ and\ \bibinfo {author} {\bibfnamefont {P.}~\bibnamefont {Grangier}},\ }\bibfield  {title} {\bibinfo {title} {Non-gaussian statistics from individual pulses of squeezed light},\ }\href {https://doi.org/10.1103/PhysRevLett.92.153601} {\bibfield  {journal} {\bibinfo  {journal} {Phys. Rev. Lett.}\ }\textbf {\bibinfo {volume} {92}},\ \bibinfo {pages} {153601} (\bibinfo {year} {2004})}\BibitemShut {NoStop}%
\bibitem [{\citenamefont {Endo}\ \emph {et~al.}(2023)\citenamefont {Endo}, \citenamefont {He}, \citenamefont {Sonoyama}, \citenamefont {Takahashi}, \citenamefont {Kashiwazaki}, \citenamefont {Umeki}, \citenamefont {Takasu}, \citenamefont {Hattori}, \citenamefont {Fukuda}, \citenamefont {Fukui}, \citenamefont {Takase}, \citenamefont {Asavanant}, \citenamefont {Marek}, \citenamefont {Filip},\ and\ \citenamefont {Furusawa}}]{Endo2023}%
  \BibitemOpen
  \bibfield  {author} {\bibinfo {author} {\bibfnamefont {M.}~\bibnamefont {Endo}}, \bibinfo {author} {\bibfnamefont {R.}~\bibnamefont {He}}, \bibinfo {author} {\bibfnamefont {T.}~\bibnamefont {Sonoyama}}, \bibinfo {author} {\bibfnamefont {K.}~\bibnamefont {Takahashi}}, \bibinfo {author} {\bibfnamefont {T.}~\bibnamefont {Kashiwazaki}}, \bibinfo {author} {\bibfnamefont {T.}~\bibnamefont {Umeki}}, \bibinfo {author} {\bibfnamefont {S.}~\bibnamefont {Takasu}}, \bibinfo {author} {\bibfnamefont {K.}~\bibnamefont {Hattori}}, \bibinfo {author} {\bibfnamefont {D.}~\bibnamefont {Fukuda}}, \bibinfo {author} {\bibfnamefont {K.}~\bibnamefont {Fukui}}, \bibinfo {author} {\bibfnamefont {K.}~\bibnamefont {Takase}}, \bibinfo {author} {\bibfnamefont {W.}~\bibnamefont {Asavanant}}, \bibinfo {author} {\bibfnamefont {P.}~\bibnamefont {Marek}}, \bibinfo {author} {\bibfnamefont {R.}~\bibnamefont {Filip}},\ and\ \bibinfo {author} {\bibfnamefont {A.}~\bibnamefont {Furusawa}},\ }\bibfield  {title} {\bibinfo {title} {Non-gaussian
  quantum state generation by multi-photon subtraction at the telecommunication wavelength},\ }\href@noop {} {\bibfield  {journal} {\bibinfo  {journal} {Opt. Express}\ }\textbf {\bibinfo {volume} {31}},\ \bibinfo {pages} {12865} (\bibinfo {year} {2023})}\BibitemShut {NoStop}%
\bibitem [{\citenamefont {Slusher}\ \emph {et~al.}(1987)\citenamefont {Slusher}, \citenamefont {Grangier}, \citenamefont {LaPorta}, \citenamefont {Yurke},\ and\ \citenamefont {Potasek}}]{Slusher1987}%
  \BibitemOpen
  \bibfield  {author} {\bibinfo {author} {\bibfnamefont {R.~E.}\ \bibnamefont {Slusher}}, \bibinfo {author} {\bibfnamefont {P.}~\bibnamefont {Grangier}}, \bibinfo {author} {\bibfnamefont {A.}~\bibnamefont {LaPorta}}, \bibinfo {author} {\bibfnamefont {B.}~\bibnamefont {Yurke}},\ and\ \bibinfo {author} {\bibfnamefont {M.~J.}\ \bibnamefont {Potasek}},\ }\bibfield  {title} {\bibinfo {title} {Pulsed squeezed light},\ }\href {https://doi.org/10.1103/PhysRevLett.59.2566} {\bibfield  {journal} {\bibinfo  {journal} {Phys. Rev. Lett.}\ }\textbf {\bibinfo {volume} {59}},\ \bibinfo {pages} {2566} (\bibinfo {year} {1987})}\BibitemShut {NoStop}%
\bibitem [{\citenamefont {Zhao}\ \emph {et~al.}(2020)\citenamefont {Zhao}, \citenamefont {Okawachi}, \citenamefont {Jang}, \citenamefont {Ji}, \citenamefont {Lipson},\ and\ \citenamefont {Gaeta}}]{Zhao2020}%
  \BibitemOpen
  \bibfield  {author} {\bibinfo {author} {\bibfnamefont {Y.}~\bibnamefont {Zhao}}, \bibinfo {author} {\bibfnamefont {Y.}~\bibnamefont {Okawachi}}, \bibinfo {author} {\bibfnamefont {J.~K.}\ \bibnamefont {Jang}}, \bibinfo {author} {\bibfnamefont {X.}~\bibnamefont {Ji}}, \bibinfo {author} {\bibfnamefont {M.}~\bibnamefont {Lipson}},\ and\ \bibinfo {author} {\bibfnamefont {A.~L.}\ \bibnamefont {Gaeta}},\ }\bibfield  {title} {\bibinfo {title} {Near-degenerate quadrature-squeezed vacuum generation on a silicon-nitride chip},\ }\href {https://doi.org/10.1103/PhysRevLett.124.193601} {\bibfield  {journal} {\bibinfo  {journal} {Phys. Rev. Lett.}\ }\textbf {\bibinfo {volume} {124}},\ \bibinfo {pages} {193601} (\bibinfo {year} {2020})}\BibitemShut {NoStop}%
\bibitem [{\citenamefont {Yang}\ \emph {et~al.}(2008)\citenamefont {Yang}, \citenamefont {Liscidini},\ and\ \citenamefont {Sipe}}]{backwardHeisenberg}%
  \BibitemOpen
  \bibfield  {author} {\bibinfo {author} {\bibfnamefont {Z.}~\bibnamefont {Yang}}, \bibinfo {author} {\bibfnamefont {M.}~\bibnamefont {Liscidini}},\ and\ \bibinfo {author} {\bibfnamefont {J.~E.}\ \bibnamefont {Sipe}},\ }\bibfield  {title} {\bibinfo {title} {Spontaneous parametric down-conversion in waveguides: A backward {H}eisenberg picture approach},\ }\href@noop {} {\bibfield  {journal} {\bibinfo  {journal} {Phys. Rev. A}\ }\textbf {\bibinfo {volume} {77}},\ \bibinfo {pages} {033808} (\bibinfo {year} {2008})}\BibitemShut {NoStop}%
\bibitem [{\citenamefont {Blanes}\ \emph {et~al.}(2009)\citenamefont {Blanes}, \citenamefont {Casas}, \citenamefont {Oteo},\ and\ \citenamefont {Ros}}]{Blanes2009}%
  \BibitemOpen
  \bibfield  {author} {\bibinfo {author} {\bibfnamefont {S.}~\bibnamefont {Blanes}}, \bibinfo {author} {\bibfnamefont {F.}~\bibnamefont {Casas}}, \bibinfo {author} {\bibfnamefont {J.}~\bibnamefont {Oteo}},\ and\ \bibinfo {author} {\bibfnamefont {J.}~\bibnamefont {Ros}},\ }\bibfield  {title} {\bibinfo {title} {The magnus expansion and some of its applications},\ }\href {https://doi.org/https://doi.org/10.1016/j.physrep.2008.11.001} {\bibfield  {journal} {\bibinfo  {journal} {Physics Reports}\ }\textbf {\bibinfo {volume} {470}},\ \bibinfo {pages} {151} (\bibinfo {year} {2009})}\BibitemShut {NoStop}%
\bibitem [{\citenamefont {Triginer}\ \emph {et~al.}(2020)\citenamefont {Triginer}, \citenamefont {Vidrighin}, \citenamefont {Quesada}, \citenamefont {Eckstein}, \citenamefont {Moore}, \citenamefont {Kolthammer}, \citenamefont {Sipe},\ and\ \citenamefont {Walmsley}}]{Triginer2020}%
  \BibitemOpen
  \bibfield  {author} {\bibinfo {author} {\bibfnamefont {G.}~\bibnamefont {Triginer}}, \bibinfo {author} {\bibfnamefont {M.~D.}\ \bibnamefont {Vidrighin}}, \bibinfo {author} {\bibfnamefont {N.}~\bibnamefont {Quesada}}, \bibinfo {author} {\bibfnamefont {A.}~\bibnamefont {Eckstein}}, \bibinfo {author} {\bibfnamefont {M.}~\bibnamefont {Moore}}, \bibinfo {author} {\bibfnamefont {W.~S.}\ \bibnamefont {Kolthammer}}, \bibinfo {author} {\bibfnamefont {J.~E.}\ \bibnamefont {Sipe}},\ and\ \bibinfo {author} {\bibfnamefont {I.~A.}\ \bibnamefont {Walmsley}},\ }\bibfield  {title} {\bibinfo {title} {Understanding high-gain twin-beam sources using cascaded stimulated emission},\ }\href@noop {} {\bibfield  {journal} {\bibinfo  {journal} {Phys. Rev. X}\ }\textbf {\bibinfo {volume} {10}},\ \bibinfo {pages} {031063} (\bibinfo {year} {2020})}\BibitemShut {NoStop}%
\bibitem [{\citenamefont {Quesada}\ \emph {et~al.}(2021)\citenamefont {Quesada}, \citenamefont {Helt}, \citenamefont {Menotti}, \citenamefont {Liscidini},\ and\ \citenamefont {Sipe}}]{beyondPhotonPairs}%
  \BibitemOpen
  \bibfield  {author} {\bibinfo {author} {\bibfnamefont {N.}~\bibnamefont {Quesada}}, \bibinfo {author} {\bibfnamefont {L.~G.}\ \bibnamefont {Helt}}, \bibinfo {author} {\bibfnamefont {M.}~\bibnamefont {Menotti}}, \bibinfo {author} {\bibfnamefont {M.}~\bibnamefont {Liscidini}},\ and\ \bibinfo {author} {\bibfnamefont {J.~E.}\ \bibnamefont {Sipe}},\ }\bibfield  {title} {\bibinfo {title} {{Beyond photon pairs: Nonlinear quantum photonics in the high-gain regime}},\ }\href@noop {} {\bibfield  {journal} {\bibinfo  {journal} {arXiv:2110.04340}\ } (\bibinfo {year} {2021})}\BibitemShut {NoStop}%
\bibitem [{\citenamefont {Quesada}\ and\ \citenamefont {Sipe}(2014)}]{quesada2014}%
  \BibitemOpen
  \bibfield  {author} {\bibinfo {author} {\bibfnamefont {N.}~\bibnamefont {Quesada}}\ and\ \bibinfo {author} {\bibfnamefont {J.~E.}\ \bibnamefont {Sipe}},\ }\bibfield  {title} {\bibinfo {title} {Effects of time ordering in quantum nonlinear optics},\ }\href {https://doi.org/10.1103/PhysRevA.90.063840} {\bibfield  {journal} {\bibinfo  {journal} {Phys. Rev. A}\ }\textbf {\bibinfo {volume} {90}},\ \bibinfo {pages} {063840} (\bibinfo {year} {2014})}\BibitemShut {NoStop}%
\bibitem [{\citenamefont {Yanagimoto}\ \emph {et~al.}(2022)\citenamefont {Yanagimoto}, \citenamefont {Ng}, \citenamefont {Yamamura}, \citenamefont {Onodera}, \citenamefont {Wright}, \citenamefont {Jankowski}, \citenamefont {Fejer}, \citenamefont {McMahon},\ and\ \citenamefont {Mabuchi}}]{Yanagimoto2022}%
  \BibitemOpen
  \bibfield  {author} {\bibinfo {author} {\bibfnamefont {R.}~\bibnamefont {Yanagimoto}}, \bibinfo {author} {\bibfnamefont {E.}~\bibnamefont {Ng}}, \bibinfo {author} {\bibfnamefont {A.}~\bibnamefont {Yamamura}}, \bibinfo {author} {\bibfnamefont {T.}~\bibnamefont {Onodera}}, \bibinfo {author} {\bibfnamefont {L.~G.}\ \bibnamefont {Wright}}, \bibinfo {author} {\bibfnamefont {M.}~\bibnamefont {Jankowski}}, \bibinfo {author} {\bibfnamefont {M.~M.}\ \bibnamefont {Fejer}}, \bibinfo {author} {\bibfnamefont {P.~L.}\ \bibnamefont {McMahon}},\ and\ \bibinfo {author} {\bibfnamefont {H.}~\bibnamefont {Mabuchi}},\ }\bibfield  {title} {\bibinfo {title} {Onset of non-gaussian quantum physics in pulsed squeezing with mesoscopic fields},\ }\href {https://doi.org/10.1364/OPTICA.447782} {\bibfield  {journal} {\bibinfo  {journal} {Optica}\ }\textbf {\bibinfo {volume} {9}},\ \bibinfo {pages} {379} (\bibinfo {year} {2022})}\BibitemShut {NoStop}%
\bibitem [{\citenamefont {Chinni}\ and\ \citenamefont {Quesada}(2023)}]{quesada2023}%
  \BibitemOpen
  \bibfield  {author} {\bibinfo {author} {\bibfnamefont {K.}~\bibnamefont {Chinni}}\ and\ \bibinfo {author} {\bibfnamefont {N.}~\bibnamefont {Quesada}},\ }\bibfield  {title} {\bibinfo {title} {Beyond the parametric approximation: pump depletion, entanglement and squeezing in macroscopic down-conversion},\ }\href@noop {} {\bibfield  {journal} {\bibinfo  {journal} {arXiv:2312.09239}\ } (\bibinfo {year} {2023})}\BibitemShut {NoStop}%
\bibitem [{\citenamefont {Fl\'{o}rez}\ \emph {et~al.}(2020)\citenamefont {Fl\'{o}rez}, \citenamefont {Lundeen},\ and\ \citenamefont {Chekhova}}]{Florez2020}%
  \BibitemOpen
  \bibfield  {author} {\bibinfo {author} {\bibfnamefont {J.}~\bibnamefont {Fl\'{o}rez}}, \bibinfo {author} {\bibfnamefont {J.~S.}\ \bibnamefont {Lundeen}},\ and\ \bibinfo {author} {\bibfnamefont {M.~V.}\ \bibnamefont {Chekhova}},\ }\bibfield  {title} {\bibinfo {title} {Pump depletion in parametric down-conversion with low pump energies},\ }\href {https://doi.org/10.1364/OL.394925} {\bibfield  {journal} {\bibinfo  {journal} {Opt. Lett.}\ }\textbf {\bibinfo {volume} {45}},\ \bibinfo {pages} {4264} (\bibinfo {year} {2020})}\BibitemShut {NoStop}%
\bibitem [{\citenamefont {Ramelow}\ \emph {et~al.}(2019)\citenamefont {Ramelow}, \citenamefont {Farsi}, \citenamefont {Vernon}, \citenamefont {Clemmen}, \citenamefont {Ji}, \citenamefont {Sipe}, \citenamefont {Liscidini}, \citenamefont {Lipson},\ and\ \citenamefont {Gaeta}}]{ramelow2019}%
  \BibitemOpen
  \bibfield  {author} {\bibinfo {author} {\bibfnamefont {S.}~\bibnamefont {Ramelow}}, \bibinfo {author} {\bibfnamefont {A.}~\bibnamefont {Farsi}}, \bibinfo {author} {\bibfnamefont {Z.}~\bibnamefont {Vernon}}, \bibinfo {author} {\bibfnamefont {S.}~\bibnamefont {Clemmen}}, \bibinfo {author} {\bibfnamefont {X.}~\bibnamefont {Ji}}, \bibinfo {author} {\bibfnamefont {J.~E.}\ \bibnamefont {Sipe}}, \bibinfo {author} {\bibfnamefont {M.}~\bibnamefont {Liscidini}}, \bibinfo {author} {\bibfnamefont {M.}~\bibnamefont {Lipson}},\ and\ \bibinfo {author} {\bibfnamefont {A.~L.}\ \bibnamefont {Gaeta}},\ }\bibfield  {title} {\bibinfo {title} {Strong nonlinear coupling in a si$_3$n$_4$ ring resonator},\ }\href {https://doi.org/10.1103/PhysRevLett.122.153906} {\bibfield  {journal} {\bibinfo  {journal} {Phys. Rev. Lett.}\ }\textbf {\bibinfo {volume} {122}},\ \bibinfo {pages} {153906} (\bibinfo {year} {2019})}\BibitemShut {NoStop}%
\bibitem [{\citenamefont {Steiner}\ \emph {et~al.}(2021)\citenamefont {Steiner}, \citenamefont {Castro}, \citenamefont {Chang}, \citenamefont {Dang}, \citenamefont {Xie}, \citenamefont {Norman}, \citenamefont {Bowers},\ and\ \citenamefont {Moody}}]{Steiner2021}%
  \BibitemOpen
  \bibfield  {author} {\bibinfo {author} {\bibfnamefont {T.~J.}\ \bibnamefont {Steiner}}, \bibinfo {author} {\bibfnamefont {J.~E.}\ \bibnamefont {Castro}}, \bibinfo {author} {\bibfnamefont {L.}~\bibnamefont {Chang}}, \bibinfo {author} {\bibfnamefont {Q.}~\bibnamefont {Dang}}, \bibinfo {author} {\bibfnamefont {W.}~\bibnamefont {Xie}}, \bibinfo {author} {\bibfnamefont {J.}~\bibnamefont {Norman}}, \bibinfo {author} {\bibfnamefont {J.~E.}\ \bibnamefont {Bowers}},\ and\ \bibinfo {author} {\bibfnamefont {G.}~\bibnamefont {Moody}},\ }\bibfield  {title} {\bibinfo {title} {Ultrabright entangled-photon-pair generation from an $\mathrm{Al}\mathrm{Ga}\mathrm{As}$-on-insulator microring resonator},\ }\href {https://doi.org/10.1103/PRXQuantum.2.010337} {\bibfield  {journal} {\bibinfo  {journal} {PRX Quantum}\ }\textbf {\bibinfo {volume} {2}},\ \bibinfo {pages} {010337} (\bibinfo {year} {2021})}\BibitemShut {NoStop}%
\bibitem [{\citenamefont {Vaidya}\ \emph {et~al.}(2020)\citenamefont {Vaidya}, \citenamefont {Morrison}, \citenamefont {Helt}, \citenamefont {Shahrokshahi}, \citenamefont {Mahler}, \citenamefont {Collins}, \citenamefont {Tan}, \citenamefont {Lavoie}, \citenamefont {Repingon}, \citenamefont {Menotti}, \citenamefont {Quesada}, \citenamefont {Pooser}, \citenamefont {Lita}, \citenamefont {Gerrits}, \citenamefont {Nam},\ and\ \citenamefont {Vernon}}]{Vaidya2020}%
  \BibitemOpen
  \bibfield  {author} {\bibinfo {author} {\bibfnamefont {V.~D.}\ \bibnamefont {Vaidya}}, \bibinfo {author} {\bibfnamefont {B.}~\bibnamefont {Morrison}}, \bibinfo {author} {\bibfnamefont {L.~G.}\ \bibnamefont {Helt}}, \bibinfo {author} {\bibfnamefont {R.}~\bibnamefont {Shahrokshahi}}, \bibinfo {author} {\bibfnamefont {D.~H.}\ \bibnamefont {Mahler}}, \bibinfo {author} {\bibfnamefont {M.~J.}\ \bibnamefont {Collins}}, \bibinfo {author} {\bibfnamefont {K.}~\bibnamefont {Tan}}, \bibinfo {author} {\bibfnamefont {J.}~\bibnamefont {Lavoie}}, \bibinfo {author} {\bibfnamefont {A.}~\bibnamefont {Repingon}}, \bibinfo {author} {\bibfnamefont {M.}~\bibnamefont {Menotti}}, \bibinfo {author} {\bibfnamefont {N.}~\bibnamefont {Quesada}}, \bibinfo {author} {\bibfnamefont {R.~C.}\ \bibnamefont {Pooser}}, \bibinfo {author} {\bibfnamefont {A.~E.}\ \bibnamefont {Lita}}, \bibinfo {author} {\bibfnamefont {T.}~\bibnamefont {Gerrits}}, \bibinfo {author} {\bibfnamefont {S.~W.}\ \bibnamefont {Nam}},\ and\ \bibinfo {author} {\bibfnamefont
  {Z.}~\bibnamefont {Vernon}},\ }\bibfield  {title} {\bibinfo {title} {Broadband quadrature-squeezed vacuum and nonclassical photon number correlations from a nanophotonic device},\ }\href@noop {} {\bibfield  {journal} {\bibinfo  {journal} {Science Advances}\ }\textbf {\bibinfo {volume} {6}},\ \bibinfo {pages} {eaba9186} (\bibinfo {year} {2020})}\BibitemShut {NoStop}%
\bibitem [{\citenamefont {Zhang}\ \emph {et~al.}(2021)\citenamefont {Zhang}, \citenamefont {Menotti}, \citenamefont {Tan}, \citenamefont {Vaidya}, \citenamefont {Mahler}, \citenamefont {Helt}, \citenamefont {Zatti}, \citenamefont {Liscidini}, \citenamefont {Morrison},\ and\ \citenamefont {Vernon}}]{Zhang2021}%
  \BibitemOpen
  \bibfield  {author} {\bibinfo {author} {\bibfnamefont {Y.}~\bibnamefont {Zhang}}, \bibinfo {author} {\bibfnamefont {M.}~\bibnamefont {Menotti}}, \bibinfo {author} {\bibfnamefont {K.}~\bibnamefont {Tan}}, \bibinfo {author} {\bibfnamefont {V.~D.}\ \bibnamefont {Vaidya}}, \bibinfo {author} {\bibfnamefont {D.~H.}\ \bibnamefont {Mahler}}, \bibinfo {author} {\bibfnamefont {L.~G.}\ \bibnamefont {Helt}}, \bibinfo {author} {\bibfnamefont {L.}~\bibnamefont {Zatti}}, \bibinfo {author} {\bibfnamefont {M.}~\bibnamefont {Liscidini}}, \bibinfo {author} {\bibfnamefont {B.}~\bibnamefont {Morrison}},\ and\ \bibinfo {author} {\bibfnamefont {Z.}~\bibnamefont {Vernon}},\ }\bibfield  {title} {\bibinfo {title} {Squeezed light from a nanophotonic molecule},\ }\href@noop {} {\bibfield  {journal} {\bibinfo  {journal} {Nat. Commun.}\ }\textbf {\bibinfo {volume} {12}},\ \bibinfo {pages} {2233} (\bibinfo {year} {2021})}\BibitemShut {NoStop}%
\bibitem [{\citenamefont {Liscidini}\ \emph {et~al.}(2012)\citenamefont {Liscidini}, \citenamefont {Helt},\ and\ \citenamefont {Sipe}}]{Liscidni2012}%
  \BibitemOpen
  \bibfield  {author} {\bibinfo {author} {\bibfnamefont {M.}~\bibnamefont {Liscidini}}, \bibinfo {author} {\bibfnamefont {L.~G.}\ \bibnamefont {Helt}},\ and\ \bibinfo {author} {\bibfnamefont {J.~E.}\ \bibnamefont {Sipe}},\ }\bibfield  {title} {\bibinfo {title} {Asymptotic fields for a hamiltonian treatment of nonlinear electromagnetic phenomena},\ }\href {https://doi.org/10.1103/PhysRevA.85.013833} {\bibfield  {journal} {\bibinfo  {journal} {Phys. Rev. A}\ }\textbf {\bibinfo {volume} {85}},\ \bibinfo {pages} {013833} (\bibinfo {year} {2012})}\BibitemShut {NoStop}%
\bibitem [{\citenamefont {Vernon}\ \emph {et~al.}(2019)\citenamefont {Vernon}, \citenamefont {Quesada}, \citenamefont {Liscidini}, \citenamefont {Morrison}, \citenamefont {Menotti}, \citenamefont {Tan},\ and\ \citenamefont {Sipe}}]{xanaduCV2019}%
  \BibitemOpen
  \bibfield  {author} {\bibinfo {author} {\bibfnamefont {Z.}~\bibnamefont {Vernon}}, \bibinfo {author} {\bibfnamefont {N.}~\bibnamefont {Quesada}}, \bibinfo {author} {\bibfnamefont {M.}~\bibnamefont {Liscidini}}, \bibinfo {author} {\bibfnamefont {B.}~\bibnamefont {Morrison}}, \bibinfo {author} {\bibfnamefont {M.}~\bibnamefont {Menotti}}, \bibinfo {author} {\bibfnamefont {K.}~\bibnamefont {Tan}},\ and\ \bibinfo {author} {\bibfnamefont {J.~E.}\ \bibnamefont {Sipe}},\ }\bibfield  {title} {\bibinfo {title} {Scalable squeezed-light source for continuous-variable quantum sampling},\ }\href@noop {} {\bibfield  {journal} {\bibinfo  {journal} {Phys. Rev. Appl.}\ }\textbf {\bibinfo {volume} {12}},\ \bibinfo {pages} {064024} (\bibinfo {year} {2019})}\BibitemShut {NoStop}%
\bibitem [{\citenamefont {Banic}\ \emph {et~al.}(2022)\citenamefont {Banic}, \citenamefont {Zatti}, \citenamefont {Liscidini},\ and\ \citenamefont {Sipe}}]{MBanicRingLoss2022}%
  \BibitemOpen
  \bibfield  {author} {\bibinfo {author} {\bibfnamefont {M.}~\bibnamefont {Banic}}, \bibinfo {author} {\bibfnamefont {L.}~\bibnamefont {Zatti}}, \bibinfo {author} {\bibfnamefont {M.}~\bibnamefont {Liscidini}},\ and\ \bibinfo {author} {\bibfnamefont {J.~E.}\ \bibnamefont {Sipe}},\ }\bibfield  {title} {\bibinfo {title} {Two strategies for modeling nonlinear optics in lossy integrated photonic structures},\ }\href {https://doi.org/10.1103/PhysRevA.106.043707} {\bibfield  {journal} {\bibinfo  {journal} {Phys. Rev. A}\ }\textbf {\bibinfo {volume} {106}},\ \bibinfo {pages} {043707} (\bibinfo {year} {2022})}\BibitemShut {NoStop}%
\bibitem [{\citenamefont {Quesada}\ \emph {et~al.}(2022)\citenamefont {Quesada}, \citenamefont {Helt}, \citenamefont {Menotti}, \citenamefont {Liscidini},\ and\ \citenamefont {Sipe}}]{Quesada2022}%
  \BibitemOpen
  \bibfield  {author} {\bibinfo {author} {\bibfnamefont {N.}~\bibnamefont {Quesada}}, \bibinfo {author} {\bibfnamefont {L.~G.}\ \bibnamefont {Helt}}, \bibinfo {author} {\bibfnamefont {M.}~\bibnamefont {Menotti}}, \bibinfo {author} {\bibfnamefont {M.}~\bibnamefont {Liscidini}},\ and\ \bibinfo {author} {\bibfnamefont {J.~E.}\ \bibnamefont {Sipe}},\ }\bibfield  {title} {\bibinfo {title} {Beyond photon pairs---nonlinear quantum photonics in the high-gain regime: a tutorial},\ }\href {https://doi.org/10.1364/AOP.445496} {\bibfield  {journal} {\bibinfo  {journal} {Adv. Opt. Photon.}\ }\textbf {\bibinfo {volume} {14}},\ \bibinfo {pages} {291} (\bibinfo {year} {2022})}\BibitemShut {NoStop}%
\bibitem [{\citenamefont {Ma}\ and\ \citenamefont {Rhodes}(1990)}]{marhodes1990}%
  \BibitemOpen
  \bibfield  {author} {\bibinfo {author} {\bibfnamefont {X.}~\bibnamefont {Ma}}\ and\ \bibinfo {author} {\bibfnamefont {W.}~\bibnamefont {Rhodes}},\ }\bibfield  {title} {\bibinfo {title} {Multimode squeeze operators and squeezed states},\ }\href@noop {} {\bibfield  {journal} {\bibinfo  {journal} {Phys. Rev. A}\ }\textbf {\bibinfo {volume} {41}},\ \bibinfo {pages} {4625} (\bibinfo {year} {1990})}\BibitemShut {NoStop}%
\bibitem [{\citenamefont {Houde}\ and\ \citenamefont {Quesada}(2023)}]{Houde2023}%
  \BibitemOpen
  \bibfield  {author} {\bibinfo {author} {\bibfnamefont {M.}~\bibnamefont {Houde}}\ and\ \bibinfo {author} {\bibfnamefont {N.}~\bibnamefont {Quesada}},\ }\bibfield  {title} {\bibinfo {title} {{Waveguided sources of consistent, single-temporal-mode squeezed light: The good, the bad, and the ugly}},\ }\href@noop {} {\bibfield  {journal} {\bibinfo  {journal} {AVS Quantum Sci.}\ }\textbf {\bibinfo {volume} {5}},\ \bibinfo {pages} {011404} (\bibinfo {year} {2023})}\BibitemShut {NoStop}%
\bibitem [{\citenamefont {Seifoory}\ \emph {et~al.}(2022)\citenamefont {Seifoory}, \citenamefont {Vernon}, \citenamefont {Mahler}, \citenamefont {Menotti}, \citenamefont {Zhang},\ and\ \citenamefont {Sipe}}]{seifoory2022}%
  \BibitemOpen
  \bibfield  {author} {\bibinfo {author} {\bibfnamefont {H.}~\bibnamefont {Seifoory}}, \bibinfo {author} {\bibfnamefont {Z.}~\bibnamefont {Vernon}}, \bibinfo {author} {\bibfnamefont {D.~H.}\ \bibnamefont {Mahler}}, \bibinfo {author} {\bibfnamefont {M.}~\bibnamefont {Menotti}}, \bibinfo {author} {\bibfnamefont {Y.}~\bibnamefont {Zhang}},\ and\ \bibinfo {author} {\bibfnamefont {J.~E.}\ \bibnamefont {Sipe}},\ }\bibfield  {title} {\bibinfo {title} {Degenerate squeezing in a dual-pumped integrated microresonator: Parasitic processes and their suppression},\ }\href {https://doi.org/10.1103/PhysRevA.105.033524} {\bibfield  {journal} {\bibinfo  {journal} {Phys. Rev. A}\ }\textbf {\bibinfo {volume} {105}},\ \bibinfo {pages} {033524} (\bibinfo {year} {2022})}\BibitemShut {NoStop}%
\bibitem [{\citenamefont {Drago}\ and\ \citenamefont {Sipe}(2023)}]{drago2023}%
  \BibitemOpen
  \bibfield  {author} {\bibinfo {author} {\bibfnamefont {C.}~\bibnamefont {Drago}}\ and\ \bibinfo {author} {\bibfnamefont {J.~E.}\ \bibnamefont {Sipe}},\ }\bibfield  {title} {\bibinfo {title} {Taking apart squeezed light},\ }\href@noop {} {\bibfield  {journal} {\bibinfo  {journal} {arXiv:2310.10919}\ } (\bibinfo {year} {2023})}\BibitemShut {NoStop}%
\end{thebibliography}%

\appendix

\section{Effective $\chi^{(2)}$ interaction in a ring resonator}
\label{sec:ring example}
In this Appendix we derive the nonlinear Hamiltonian for an effective $\chi^{(2)}$ interaction in a ring resonator. 

We begin with the nonlinear Hamiltonian for SFWM in Eq. \eqref{eq:HSFWM}. In the interaction picture it takes the form
\begin{align}
\label{eq:Hint}
    \overline{H}_{NL}(t) & = -\frac{3}{\epsilon_0}\sum_{n,n'}\int dk_1dk_2dk_3dk_4 \nonumber
    \\
    &\times K^{nn'}(k_1,k_2,k_3,k_4) {\rm e}^{-i\Omega_{nn'}(k_1,k_2,k_3,k_4)t} \nonumber
    \\
    &\times a^\dagger_{n S k_1}a^\dagger_{n' S k_2} b_{P k_3}b_{C k_4} + {\rm H.c.},
\end{align}
where we have used the definition in Eq. \eqref{eq:interaction H}, and 
defined
\begin{align}
    \Omega_{nn'}(k_1,k_2,k_3,k_4) \equiv \omega_{Ck_4}+ \omega_{Pk_3} - \omega_{n'Sk_2}-\omega_{nSk_1}.
\end{align}
In order to build the effective $\chi^{(2)}$ interaction, the pump in the mode $C$ is treated classically with a cw field that is  undepleted. 
The  operators $b_{Ck_4}$ are replaced with the amplitudes \cite{MBanicRingLoss2022}
\begin{align}
    \label{eq:classical amplitude c.w.}
    b_{Ck_4} \rightarrow \sqrt{\frac{2\pi P_C}{\hbar \omega_C v_C}}\delta(k_4 - K_C),
\end{align}
where $P_C$ is the cw power and $v_C$ is the group velocity of the cw pump in the actual input channel.
Under this approximation, Eq. \eqref{eq:Hint}  becomes
\begin{align}
\label{eq:Hint spdc}
    \overline{H}_{NL}(t) & = \hbar \sum_{n,n'}\int dk_1 dk_2 dk_3\Lambda^{nn'}(k_1,k_2,k_3,t)  \nonumber
    \\
    &\times a^\dagger_{n S k_1}a^\dagger_{n' S k_2}b_{Pk_3}+ {\rm H.c.},
\end{align}
 where the effective $\chi^{(2)}$  coefficient  $\Lambda^{nn'}(k_1,k_2,k_3,t)$ is given by
\begin{align}
\label{eq:lambda}
   \Lambda^{nn'}(k_1,k_2,k_3,t) = &-\frac{3}{\hbar\epsilon_0 }\sqrt{\frac{2\pi P_C}{\hbar \omega_C v_C}}K^{nn'}(k_1,k_2,k_3,K_C)\nonumber
\\
   &\times  {\rm e}^{-i\Omega_{nn'}(k_1,k_2,k_3,K_C)t} .
\end{align}

We consider a ring resonator that is point-coupled to an actual waveguide that couples the light out of the ring and a phantom waveguide that handles the scattering loss; recall Fig. \ref{fig:ring}(b). The circumference of the ring is given by $\mathcal{L} = 2\pi R$, where $R$ is its radius. The resonant wavenumbers for the ring are given by $\kappa$, such that $\kappa \mathcal{L}  =2\pi m$, where $m$ is a positive integer. We assume that around each ring resonance we can neglect group velocity dispersion, and that the dispersion relation is the same for the actual and phantom waveguides. Under these assumptions the dispersion relation for either channel is given by
\begin{align}
    \label{eq:linear dispersion}
    \omega_J(k) = \omega_J + v_{J}(k - K_J),
\end{align}
where $\omega_J$ is the center frequency of the resonance $J$, $v_J$ is the group velocity in either channel, and $K_J$ is the wavenumber for the light in either channel with frequency $\omega_J$. The linewidth of the ring resonances are given by $2\overline{\Gamma}_{J}$, which includes coupling losses to the actual waveguide and scattering losses to the phantom waveguide; the loaded quality factor is then $Q^{load}_J = \omega_J/(2\overline{\Gamma}_J)$.

The asymptotic -in and -out displacement fields are defined over all space, but we assume that the nonlinear interaction takes place mostly inside the ring resonator.  Thus to obtain the nonlinear coefficient we only require expressions for the field that hold at positions in the ring. To simplify the integration of that region of space, we introduce the coordinates of the ring-frame as $(\bm r_{\perp}, \zeta)$, where $\bm r_\perp = (\rho, z)$ and $\zeta = \phi R$, and $(\rho, \phi, z)$ are cylindrical coordinates. 
For positions inside the ring, the asymptotic-in pump fields can be written as \cite{MBanicRingLoss2022}
\begin{align}
\label{eq:DP ring}
\bm D^{\textit{in}}_{P  k}(\bm r) &= -\sqrt{\frac{\hbar \omega_{P}}{4\pi}}\bm d_{P}(\bm r_{\perp};\zeta){\rm e}^{i \zeta \kappa_{P}}F_{{P}-}(k) ,
\\
\label{eq:DC ring}
\bm D^{\textit{in}}_{C  k}(\bm r) &= -\sqrt{\frac{\hbar \omega_{C}}{4\pi}}\bm d_{C}(\bm r_{\perp};\zeta){\rm e}^{i \zeta \kappa_{C}}F_{{C}-}(k) ,
\end{align}
and the asymptotic-out fields can be written as
\begin{align}
\label{eq:DS ring}
\bm D^{\textit{out}}_{n S k}(\bm r) &= -\sqrt{\frac{\hbar \omega_{S}}{4\pi}}\bm d_{S}(\bm r_{\perp};\zeta){\rm e}^{i \zeta \kappa_{S}}F^{(n)}_{{S}+}(k).
\end{align}
Here we introduced the field enhancement factors for the actual input channel, $ F_{J -}(k)$, given by \cite{MBanicRingLoss2022}
\begin{align}
\label{eq:field enhance in}
 F_{J -}(k) =\frac{1}{\sqrt{\mathcal{L}}}\left( \frac{\left(\gamma_{J}\right)^*}{v_{J}(K_{J} - k)- i\overline{\Gamma}_{J}}\right)  ,
 \end{align}
where $J=P,C$ and $\gamma_J$ is a coupling constant between a discrete ring mode and the continuous waveguide mode of the actual input channel at the frequency band $J$. The field enhancement factors for the output channels, $ F^{(n)}_{S +}(k)$, are given by
\begin{align}
\label{eq:field enhance out}
 F^{(n)}_{S+}(k) =\frac{1}{\sqrt{\mathcal{L}}}\left( \frac{\left(\gamma^{(n)}_{S}\right)^*}{v^{(n)}_{S}(K^{(n)}_{S} - k)+ i\overline{\Gamma}_{S}}\right)  ,
 \end{align}
where now the coupling constant, group velocity, and center wavenumber depend on the output channel $n$. The coupling constant for a given channel is related to the intensity decay rate into that channel, as introduced earlier \cite{MBanicRingLoss2022}. For simplicity we assume that the coupling constants are real. 

Putting Eqs. \eqref{eq:DP ring}, \eqref{eq:DC ring}, and \eqref{eq:DS ring} into Eq. \eqref{eq:K}, the nonlinear coefficient for SFWM can be written as
    \begin{align}
\label{eq:K ring}
K^{nn'}(k_1,k_2,k_3,k_4) &=\frac{\hbar^2 \epsilon_0 v_{P}v_{C}}{12\pi^2}\overline{\gamma}_{nl}\omega_S\mathcal{L} F^{(n)*}_{{S}+}(k_1)  \nonumber
\\
&\times F^{(n')*}_{{S}+}(k_2) F_{{P}-}(k_3)F_{{C}-}(k_4), 
\end{align}
where we have introduced the nonlinear parameter \cite{MBanicRingLoss2022}
\begin{align}
\label{eq:gammanl SFWM}
\overline{\gamma}_{nl} &=  \frac{3\sqrt{\omega_{P} \omega_{C}}}{4 \epsilon_0 v_{P}v_{C}\mathcal{L}}\int_{{\rm ring}} d r_\perp d\zeta \Gamma^{ijkl}_{(3)}(\bm r)\nonumber
\\
&\times d^{*i}_{S}(\bm r_{\perp};\zeta) d^{*j}_{S}(\bm r_{\perp};\zeta)  d^k_{P}(\bm r_{\perp};\zeta)  d^l_{C}(\bm r_{\perp};\zeta) {\rm e}^{i \Delta \kappa \zeta },
\end{align}
which has units of $({\rm Wm})^{-1}$, and  $\Delta k = \kappa_{P} + \kappa_C - 2\kappa_S$ is the phase matching function. 

Now to obtain an effective second-order nonlinear coefficient in the ring, we put Eq. \eqref{eq:K ring} into Eq. \eqref{eq:lambda}, and find
\begin{align}
\label{eq:lambda ring}
   \Lambda^{nn'}(k_1,k_2,k_3,t)&=-\frac{\hbar v_Pv_C\overline{\gamma}_{nl}\omega_S \mathcal{L}}{4\pi^2 } \sqrt{\frac{2\pi P_C}{\hbar \omega_C v_C}} \nonumber
   \\
   &\times F^{(n)*}_{{S}+}(k_1) F^{(n')*}_{{S}+}(k_2)F_{{P}-}(k_3)  \nonumber
   \\
   &\times{F_{{C}-}(K_C)\rm e}^{-i\Omega_{nn'}(k_1,k_2,k_3,K_C)t}.
\end{align}
Then putting Eq. \eqref{eq:lambda ring} into Eq. \eqref{eq:Hint spdc} we obtain the nonlinear Hamiltonian for an effective $\chi^{(2)}$ interaction in a ring resonator.  In order to work with this Hamiltonian numerically the wavenumbers in all channels have to be discretized.   Each waveguide mode has a discrete label $j$, such that its wavenumber is given by $k_j = 2\pi m_j/L$, where $m_j$ is an integer mode number, and $L$ is the quantization length \cite{Quesada2022}. Using this discrete notation, the nonlinear Hamiltonian in Eq. \eqref{eq:Hint spdc}  can be written as
\begin{align}
\label{eq:Hint spdc discrete}
\overline{H}_{NL}(t)&=  \hbar \sum_{n,n'}\sum_{i,j,l}\Lambda^{nn'}_{ijl}(t)a^\dagger_{nSi}a^\dagger_{n'Sj} b_{Pl} + {\rm H.c.},
\end{align}
where we defined
\begin{align}
\label{eq:discrete Lambda}
    \Lambda^{nn'}_{ijl}(t) \equiv  \left(\frac{2\pi}{L}\right)^{3/2}\Lambda^{nn'}(k_{1i},k_{2j},k_{3l},t),
\end{align}
where $\Lambda^{nn'}(k_{1i},k_{2j},k_{3l},t)$ is the effective second-order nonlinear coefficient for the ring given in Eq. \eqref{eq:gammanl SFWM},
and the operators satisfy the commutation relations
\begin{align}
 \label{eq:commute discrete}
 [a_{n J i}, a^\dagger_{n J j}]  = [b_{P i}, b^\dagger_{ P j}] = \delta_{ij}.
 \end{align}
The Hamiltonian in Eq. \eqref{eq:Hint spdc discrete}  can be written equivalently as
\begin{align}
\label{eq:Hint spdc discrete 2}
    \overline{H}_{NL}(t)&=  \hbar \sum_{(i,n),(j,n')}\sum_{l}\Lambda_{(i,n)(j,n')l}(t)\nonumber
    \\
    &\times a^\dagger_{S(i,n)}a^\dagger_{S(j,n')} b_{Pl} + {\rm H.c.}.
\end{align}
We then introduce the discrete indices $\mu$ and $\nu$ that are constructed by grouping the labels $i$ and $n$ and $j$ and $n'$ together as
\begin{align}
    \mu &\equiv (i,n),
    \\
    \nu &\equiv (j,n'),
\end{align}
where  $\mu$ refers to a photon in the output channel $n$ with wavenumber $k_i$. With this labelling we obtain the Hamiltonian in Eq. \eqref{eq:Hint spdc discrete 3}.


\section{Deriving equations for the Gaussian parameters}
\label{sec:Heff simp}
In this Appendix we show how the Eq. \eqref{eq:Uadot} for $U_a(t)$ leads to the Eqs. \eqref{eq:Vdot} and \eqref{eq:Wdot} for $\bm V(t)$ and $\bm W(t)$, and how the Eq. \eqref{eq:Ubdot} for $U_b(t)$ leads to the Eqs. \eqref{eq:dot beta} and \eqref{eq:dot thetab} for $\bm \beta(t)$ and $\theta_b(t)$.

Beginning with Eq. \eqref{eq:Uadot} for $U_a(t)$, it was pointed out earlier
\cite{marhodes1990} that, with an $H_a(t)$ of the form \eqref{eq:Ha},     its unique solution can indeed be written as 
\eqref{eq:Ua}. The task here is just to identify how the parameters in $U_a(t)$ depend on time for a specified $H_a(t)$. The strategy that we employ relies on introducing time-dependent output operators $\bm a(t)$ defined as
\begin{align}
\label{eq:a(t)}
    \bm a(t) &= U^\dagger_a(t) \bm a U_a(t) \nonumber
    \\
    &= \bm V(t) \bm a + \bm W(t) \bm a^\dagger,
\end{align}
where $\bm V(t)$ and $\bm W(t)$ are found to be given by
\begin{align}
\label{eq:Vmat}
\bm V(t) &\equiv \cosh(\bm u(t)){\rm e}^{i\bm \phi(t)} ,
\\
\label{eq:Wmat}
\bm W(t)&\equiv \sinh(\bm u(t)){\rm e}^{i\bm \alpha(t)}{\rm e}^{-i\bm \phi^{\rm T}(t)},
\end{align}
and where the Hermitian matrices $\bm u(t)$ and $\bm \alpha(t)$ that appear here come from the polar decomposition of the squeezing matrix, 
\begin{align}
\label{eq:polarJ}
    \bm J(t) = \bm u(t) \exp(i\bm \alpha(t)).
\end{align} 
The matrices $\bm V(t)$ and $\bm W(t)$  in Eqs. \eqref{eq:Vmat} and \eqref{eq:Wmat} satisfy the constraints
\begin{align}
\bm V(t) \bm V^\dagger(t) - \bm W(t)\bm W^\dagger(t) &= I,
\\
\bm V(t) \bm W^{\rm T}(t) - \bm W(t) \bm V^{\rm T}(t)&=0,
\end{align}
which are necessary so that the operators in Eq. \eqref{eq:a(t)} satisfy the usual equal time bosonic commutation relations.
Taking the time-derivative of the first line in Eq. \eqref{eq:a(t)} we obtain the  equation of motion
\begin{align}
\label{eq:H eq motion}
    \frac{d \bm a(t)}{dt} = \frac{i}{\hbar}U_a^\dagger(t)[H_a(t) , \bm a]U_a(t),
\end{align}
where we used Eq. \eqref{eq:Uadot}. Now using  Eqs. \eqref{eq:Ha} and \eqref{eq:a(t)}, Eq. \eqref{eq:H eq motion} becomes
\begin{align}
\label{eq:H eq motion 2}
    \frac{d \bm a(t)}{dt} &=-2i\bm \zeta(t)\bm W^*(t) \bm a - 2i\bm \zeta(t)\bm V^*(t) \bm a^\dagger,
\end{align}
where  $\bm \zeta(t)$ is given by Eq. \eqref{eq:zeta}. On the other hand, taking the time derivative of the second line in Eq. \eqref{eq:a(t)}, we have equivalently
\begin{align}
\label{eq:H eq motion 3}
    \frac{d \bm a(t)}{dt} &=\frac{d\bm V(t)}{dt}\bm a+\frac{d\bm W(t)}{dt}\bm a^\dagger.
\end{align}
Since the expressions for $d\bm a(t)/dt$ in Eq. \eqref{eq:H eq motion 3} and Eq. \eqref{eq:H eq motion 2} are equivalent, we can equate the elements of the matrix coefficients multiplying the $\bm a$ and $\bm a^\dagger$ operators in each expression, leading to the coupled equations Eqs. \eqref{eq:Vdot} and \eqref{eq:Wdot}.

Next we turn to deriving Eq. \eqref{eq:dot beta} for $\bm \beta(t)$ and Eq. \eqref{eq:dot thetab} for $\theta_b(t)$. We start by multiplying  Eq. \eqref{eq:Ubdot} by $U^\dagger_b(t)$ from the left
\begin{align}
    \label{eq:Ubdot 2}
    i\hbar U^\dagger_b\frac{dU_b(t)}{dt}  &= \hbar [\bm \gamma(t)]_l b^\dagger_l + {\rm H.c.},
\end{align}
where $\bm \gamma(t)$ is given by Eq. \eqref{eq:gamma(t)}. Putting Eq. \eqref{eq:Ub} into the left hand side of Eq. \eqref{eq:Ubdot 2} we obtain
\begin{align}
    \label{eq:Ubdot 3}
    i\hbar D^\dagger_b\frac{dD_b(t)}{dt} - \hbar \frac{d\theta_b(t)}{dt}  &= \hbar [\bm \gamma(t)]_l b^\dagger_l + {\rm H.c.}.
\end{align}
The derivative of the displacement operator is obtained by disentangling it and using the chain rule. Doing this we obtain
\begin{align}
\label{eq:Db dot}
    i\hbar D^\dagger_b\frac{dD_b(t)}{dt}&=i\hbar\frac{d[{\bm \beta}(t)]_l}{dt}b^\dagger_l    +\frac{i}{2}\hbar\frac{d[{\bm \beta}(t)]_l}{dt}[{\bm \beta}^*(t)]_l+{\rm H.c.}.
\end{align}
Putting Eq. \eqref{eq:Db dot} into Eq. \eqref{eq:Ubdot 3} and collecting terms, we obtain
\begin{align}
\label{eq:Ubdot 4}
    0&=\left(i\frac{d[{\bm \beta}(t)]_l}{dt}  - [\bm \gamma(t)]_l \right)b^\dagger_l    - \left(i\frac{d[{\bm \beta}^*(t)]_l}{dt}  + [\bm \gamma^*(t)]_l \right)b_l \nonumber
    \\
    &+\frac{i}{2}\left(\frac{d[{\bm \beta}(t)]_l}{dt}[{\bm \beta}^*(t)]_l - \frac{d[{\bm \beta}^*(t)]_l}{dt}[{\bm \beta}(t)]_l \right)- \frac{d\theta_b(t)}{dt},
\end{align}
which leads to
Eqs. \eqref{eq:dot beta} and \eqref{eq:dot thetab}.


\section{Extracting $\bm J(t)$ and $\bm \phi(t)$ from $\bm V(t)$ and $\bm W(t)$}
\label{sec:extract J}
We begin by doing a numerical polar decomposition of $\bm V(t)$. Note that Eq. \eqref{eq:Vmat} is already in the form of a polar decomposition, where  $\cosh(\bm u(t))$ is a Hermitian positive semi-definite (PSD) matrix and  ${\rm e}^{i\bm \phi(t)}$ is a unitary matrix. And since the Hermitian PSD matrix in a polar decomposition is always unique, the diagonal elements of $\cosh(\bm u(t))$, which we identify by the set $\{s_i(t) \}$, will be unique.  Further, since the matrices $\cosh(\bm u(t))$ and $\bm u(t)$ commute they can be diagonalized by the same unitary transformation. So we can write 
\begin{align}
\label{eq:coshdecomp}
 \cosh (\bm u(t)) = \bm F(t) {\rm diag}(s_1(t), \ldots) \bm F^{\dagger}(t).
\end{align}
and
\begin{align}
\label{eq:udiag}
   \bm u(t) =\bm F(t) 
   {\rm diag}(r_1(t),\ldots) \bm F^{\dagger}(t),    
\end{align}
where the unitary matrix $\bm F(t)$ is the same in both decompositions, and the set $\{ r_i(t) \}$ are the diagonal elements of $\bm u(t)$. Since Eqs. \eqref{eq:coshdecomp} and \eqref{eq:udiag} are diagonalizations of PSD Hermitian matrices, we have 
\begin{align}
\label {eq:sPSD}
s_i(t) \ge 0
\end{align}
and
\begin{align}
\label {eq:rPSD}
r_i(t) \ge 0.
\end{align}
But a condition stronger than Eq. \eqref{eq:sPSD} holds. Writing
Eq. \eqref{eq:udiag}  as 
\begin{align}
    \label{eq:udiagschematic}
    \bm u(t) = \bm F(t) \bm r(t) \bm F^{\dagger}(t),
\end{align}
where $\bm r(t)$ is a diagonal matrix we have 
\begin{align}
    \cosh(\bm u(t)) &= \cosh(\bm F(t) \bm r(t) \bm F^{\dagger}(t)) \nonumber
    \\
    &= \bm F(t) \cosh(\bm r(t)) \bm F^{\dagger}(t),
\end{align}
or 
\begin{align}
\label{eq:Fdetermine}
    \cosh (\bm u(t)) = \bm F(t) {\rm diag}(\cosh r_1(t),\ldots) \bm F^{\dagger}(t).
\end{align}
Comparing with Eq. \eqref{eq:coshdecomp} we see that 
\begin{align}
\label{eq:sr}
    s_i(t) = \cosh(r_i(t)), 
\end{align}
and so, given Eq. \eqref{eq:rPSD}, we have 
\begin{align}
    s_i(t) \ge 1.
\end{align}

A first consequence of this, which follows from using it in Eq. \eqref{eq:coshdecomp},
is that $\cosh (\bm u(t))$ is 
invertible and then so is $\bm V(t)$; thus from the numerical
polar decomposition of $\bm V(t)$
we not only extract a unique Hermitian matrix $\cosh (\bm u(t))$ but also a
unique unitary matrix ${\rm e}^{i \bm \phi(t)}$. A second consequence is that from Eq. \eqref{eq:sr} we can write Eq. \eqref{eq:udiag} as
\begin{align}
\label{eq:uresult}
    \bm u(t)= \bm F(t) {\rm diag}
    ( {\rm arccosh} (s_1(t)), \ldots) \bm F^{\dagger}(t)
\end{align}
where by ``${\rm arccosh}$" we mean the non-negative ${\rm arccosh}$, by virtue of Eq. \eqref{eq:rPSD}. 
In summary, then, from the numerical polar decomposition of $\bm V(t)$ we immediately obtain the unique matrices ${\rm e}^{i \bm \phi(t)}$ and $\cosh (\bm u(t))$, from the diagonalization Eq. \eqref{eq:coshdecomp} we determine the unitary matrix $\bm F(t)$ and the $\{ s_i(t) \}$, and then from Eq. \eqref{eq:uresult} we construct $\bm u(t)$.

With $\bm u(t)$ in hand we can then form 
\begin{align}
    \sinh (\bm u(t)) &= \sinh (\bm F(t) \bm r(t) \bm F^{\dagger}(t)) \nonumber
    \\
    &= \bm F(t) \sinh (\bm r(t)) \bm F^{\dagger}(t), 
\end{align}
or equivalently 
\begin{align}
\label{eq:sinhu}
    \sinh(\bm u(t)) = \bm F(t) {\rm diag}(\sinh(r_1(t)),\ldots) \bm F^{\dagger}(t),
\end{align}
which can be written as
\begin{align}
    \sinh (\bm u(t)) &= \bm F(t) {\rm diag}\left(\frac{\sinh (r_1(t))}{r_1(t)},\ldots \right) \nonumber
    \\
    &\times {\rm diag}(r_1(t),\ldots) \bm F^{\dagger}(t),
\end{align}
or
\begin{align}
\label{eq:sinhwork}
    \sinh (\bm u(t)) = \bm K(t) \bm u(t),
\end{align}
where we have used Eq. \eqref{eq:udiag} and let
\begin{align}
    \bm K(t) = \bm F(t) {\rm diag}\left(\frac{\sinh (r_1(t))}{r_1(t)},\dots\right) \bm F^{\dagger}(t),
\end{align}
with 
\begin{align}
\label{eq:Kdef}
    K_i(t) = \frac{\sinh (r_i(t))}{r_i(t)}
\end{align}
Forming $K_i(t)$ seems problematic if $r_i(t)=0$. Of course, in such an instance the natural choice would be to take $K_i(t)$ to be the limit of Eq. \eqref{eq:Kdef} as $r_i(t) \rightarrow 0$, which would give $K_i(t) = 1$. But one could worry about issues of uniqueness.  In fact, we show below that if $r_i(t)=0$ we can take $K_i(t)$ to be any nonzero number, and the final result of the following calculation of $\bm J(t)$ will be the same.

The matrix $\bm K(t)$ is clearly invertible, since we can identify 
\begin{align}
\label{eq:Kinv}
   \bm K^{-1}(t) = \bm F(t) {\rm diag}\left(\frac{1}{K_1(t)},\ldots \right)\bm F^{\dagger}(t), 
\end{align}
and furthermore we can numerically construct $\bm K^{-1}(t)$ since we know $\bm F(t)$ and the $r_i(t)$. From Eq. \eqref{eq:sinhwork} we then have 
\begin{align}
\label{eq:Kinv}
    \bm K^{-1}(t) \sinh(\bm u(t)) = \bm u(t),
\end{align}
and from Eq. \eqref{eq:Wmat}
we have
\begin{align}
     \bm K^{-1}(t) \bm W(t) &=  \bm u(t) {\rm e}^{i\bm \alpha(t)}{\rm e}^{-i\bm \phi^{\rm T}(t)} = \bm J(t) {\rm e}^{-i\bm \phi^{\rm T}(t)},  
\end{align}
so
\begin{align}
\label{eq:J matrix}
    \bm J(t) = \bm K^{-1}(t) \bm W(t) {\rm e}^{i\bm \phi^{\rm T}(t)}. 
\end{align}
All the quantities on the right-hand-side are known, and so we can determine $\bm J(t)$.

To investigate what happens to $\bm J(t)$ when some of the eigenvalues $r_i(t)$ are zero, consider
\begin{align}
    \label{eq:Kinv W}
    \bm K^{-1}(t) \bm W(t)  &= \bm F(t) {\rm diag}\left(\frac{\sinh (r_1(t))}{K_1(t)} ,\ldots \right) \nonumber
    \\
    &\times \bm F^\dagger(t){\rm e}^{i\bm \alpha(t)}{\rm e}^{-i\bm \phi^{\rm T}(t)},
\end{align}
where we used Eqs. \eqref{eq:Kinv} and \eqref{eq:sinhu}.
Now we can see that for any $i$ for which $r_i(t)=0$ we will have $\sinh(r_i(t))/K_i(t)=0$, regardless of which nonzero value $K_i(t)$ is set. 
So the eigenvalues $r_i(t)$ that are zero do not contribute to the matrix $\bm J(t)$.

At this point we have identified $\bm J(t)$ and ${\rm e}^{i\bm \phi(t)}$; we still must extract $\bm \phi(t)$. The $\bm \phi(t)$ appearing in $R_a(\bm \phi(t))$ of \eqref{eq:R} is a Hermitian matrix, and so can be diagonalized,
\begin{align}
\label{eq:phiguess}
    \bm \phi(t)= \bm M(t) {\rm diag} (\chi_1(t),\dots) \bm M^{\dagger}(t),
\end{align}
where $\bm M(t)$ is a unitary matrix. Thus we have 
\begin{align}
    {\rm e}^{i \bm \phi(t)} &={\rm e}^{i \bm M(t) {\rm diag} (\chi_1(t),\ldots) \bm M^{\dagger}(t) } \nonumber
    \\
    &= \bm M(t){\rm e}^{i {\rm diag} (\chi_1(t),\ldots)}\bm M^{\dagger}(t).   
\end{align}
And so numerically diagonalizing the identified ${\rm e}^{i \bm \phi(t)}$ would naturally lead us to identify $\bm \phi(t)$ as given by Eq. \eqref{eq:phiguess}. But although the eigenvalues of ${\rm e}^{i \bm \phi(t)}$ are well-defined, the values of the phases appearing in those eigenvalues are not. For we could just as easily write \begin{align}
  {\rm e}^{i \bm \phi(t)} &= \bm M(t){\rm e}^{i {\rm diag} (\chi_1(t)+2\pi m_1(t),\ldots)}\bm M^{\dagger}(t) \nonumber
  \\
  &= {\rm e}^{i \bm M(t) {\rm diag} (\chi_1(t)+2 \pi m_1(t),\ldots) \bm M^{\dagger}(t) }, 
\end{align}
where $m_1(t)$, etc., are any integers.  So instead of Eq. \eqref{eq:phiguess} we could just as well claim that $\bm \phi(t)$
is given by 
\begin{align}
\label{eq:phiguess2}
     \bm \phi(t)&= \bm M(t) {\rm diag} (\chi_1(t)+ 2 \pi m_1(t),...) \bm M^{\dagger}(t).  
\end{align}
However, this ambiguity is benign, because both Eq. \eqref{eq:phiguess} and Eq. \eqref{eq:phiguess2} lead to the same effective $R_a(\bm \phi(t))$.

To see this, note that if we posit Eq. \eqref{eq:phiguess2} we have 
\begin{align}
  [\bm \phi(t)]_{kl}=  [\bm M(t)]_{kp}(\chi_p(t)+2 \pi m_p(t)) [\bm M^{\dagger}(t)]_{pl},  
\end{align}
and so 
\begin{align}
\label{eq:Rwork}
    R_a(\bm \phi(t)) = {\rm e}^{i  a^{\dagger}_k [\bm M(t)]_{kp} (\chi_p(t) +2 \pi m_p(t)) [\bm M^{\dagger}(t)]_{pl} a_l}.
\end{align}
Now put 
\begin{align}
    c_p(t) = [\bm M^{\dagger}(t)]_{pl} a_l = [\bm M^{*}(t)]_{lp} a_l,
\end{align}
so
\begin{align}
    c^{\dagger}_q(t)= [\bm M(t)]_{jq} a^{\dagger}_j. 
\end{align}
Thus
\begin{align}
    [c_p(t),c^{\dagger}_q(t)] = \delta_{pq},
\end{align}
and the new operators $c_p(t)$ and $c^{\dagger}_q(t)$ satisfy equal time commutation relations of raising and lowering harmonic oscillator operators.  In terms of them, we can write \eqref{eq:Rwork} as
\begin{align}
\label{eq:Rwork2}
    R_a(\bm \phi(t)) &= {\rm e}^{i  c^{\dagger}_p(t)(\chi_p(t)+2 \pi m_p(t)) c_p(t)} \nonumber
    \\
    &= {\rm e}^{i c^{\dagger}_p(t) \chi_p(t) c_p(t)} {\rm e}^ {2 \pi i  m_p(t) c^{\dagger}_p(t) c_p(t)}.
\end{align}
As basis kets we can choose the eigenkets $| \{n\},t \rangle$ of all the number operators $c^{\dagger}_q(t)c_q(t)$,
where 
\begin{align}
    c^{\dagger}_q(t) c_q(t) | \{ n \},t \rangle  = n_q| \{n \},t \rangle. 
\end{align} Then
\begin{align}
\label{eq:numberwork}
  {\rm e}^ {2 \pi i m_p(t) c^{\dagger}_p(t) c_p(t)} | \{ n \} ,t \rangle = {\rm e}^{2 \pi i  m_p(t) n_p } | \{ n \} ,t \rangle = | \{ n \} ,t \rangle,
\end{align}
and since any ket $ | \varphi(t) \rangle $ can be written as a superposition of the $| \{ n \} ,t \rangle $, from \eqref{eq:Rwork2} and \eqref{eq:numberwork} we have 
\begin{align}
    R_a(\bm \phi(t)) | \varphi(t) \rangle &= {\rm e}^{i  c^{\dagger}_p(t)(\chi_p(t)+2 \pi m_p(t)) c_p(t)} | \varphi(t) \rangle \nonumber
    \\
    &= {\rm e}^{i c^{\dagger}_p(t) \chi_p(t) c_p(t)} | \varphi(t) \rangle.
\end{align}
Thus the result we find for $R_a(\bm \phi(t))$ will be the same whether we use  Eq. \eqref{eq:phiguess} or Eq. \eqref{eq:phiguess2} in identifying the logarithm of ${\rm e}^{i \bm \phi(t)}$ to construct $R_a(\bm \phi(t))$; any logarithm of ${\rm e}^{i \bm \phi(t)}$ will do.


\section{Determining the phase $\theta_a(t)$}
\label{sec:theta}
The phase $\theta_a(t)$ appears in the unitary operator $U_a(t)$ in Eq. \eqref{eq:Ua}. In this section we determine this phase following the approach of Ma and Rhodes \cite{marhodes1990}. Recall that $U_a(t)$ is the solution to the Schr\"odinger equation in Eq. \eqref{eq:Uadot} with the Hamiltonian $H_a(t)$ defined in Eq. \eqref{eq:Ha}. Given the form of $H_a(t)$ in Eq. \eqref{eq:Ha}, it follows \cite{marhodes1990} that the phase $\theta_a(t)$ is given by
\begin{align}
\label{eq:phase a appendix}
    \theta_a(t) = -\frac{1}{2} \int_{-\infty}^t dt' {\rm tr}\left(\bm \zeta^*(t') \tanh(\bm u(t')){\rm e}^{i\bm \alpha(t')}\right) +{\rm c.c.},
\end{align}
where $\bm \zeta(t)$ is defined in Eq. \eqref{eq:zeta} and is related to the coefficients in $H_a(t)$; recall Eq. \eqref{eq:polarJ} for the polar decomposition of the squeezing matrix $\bm J(t) = \bm u(t) \exp(i\bm \alpha(t))$.

The first-order solution that is discussed in Sec. \ref{sec:low} is valid when the amount of squeezing is small or the nonlinearity is weak.  In this subsection we show that the phase $\theta_a(t)$ in Eq. \eqref{eq:phase a appendix} is approximately zero in the first-order solution. In this case because the entries of the squeezing matrix $\bm J(t)$ are small  we replace $\tanh(\bm u(t'))\exp(i\bm \alpha(t'))$ in Eq. \eqref{eq:phase a appendix} with $\bm u(t')\exp(i\bm \alpha(t')) = \bm J(t')$ (recall Eq. \eqref{eq:polarJ}). Doing this we can write Eq. \eqref{eq:phase a appendix} approximately as
\begin{align}
\label{eq:phase a appendix approx}
    \theta_a(t) \approx -\frac{1}{2} \int_{-\infty}^t dt' {\rm tr}\left(\bm \zeta^*(t') \bm J(t')\right) +{\rm c.c.}.
\end{align}
But from Eq. \eqref{eq:J lowgain} we have that $\bm J(t') = -2i\int_{-\infty}^{t'} dt''\bm \zeta(t'')$. Putting this into Eq. \eqref{eq:phase a appendix approx}, the term inside the trace is $\bm \zeta^*(t')\bm \zeta(t'')$, which is much smaller than unity. Thus $\theta_a(t)$ is approximately zero. 

\section{A conserved quantity}
\label{sec:conservedQ}
We define the operator $\hat{Q}$ 
\begin{align}
\label{eq:Qop}
\hat{Q} \equiv \frac{1}{2}{\bm a^\dagger}^{\rm T} \bm a + {\bm b^\dagger}^{\rm T} \bm b,
\end{align}
which is a conserved quantity with respect to the full Hamiltonian in Eq. \eqref{eq:Hint spdc discrete 3}, that is 
\begin{align}
\label{eq:Qcons}
[\hat{Q},\overline{H}_{NL}(t)] = 0.
\end{align}
But $\hat{Q}$ is not a conserved quantity with respect to the effective Hamiltonian in Eq. \eqref{eq:HI}. In this Appendix we show that the average value of $\hat{Q}$ in the Gaussian limit is nonetheless a conserved quantity.

We begin with expectation value of $\hat{Q}$ using the ket in the Gaussian limit given by Eq. \eqref{eq:Gaussian ket} 
\begin{align}
\label{eq:Q}
\langle \hat{Q}\rangle &\equiv \bra{\overline{\psi}_G(t)} \hat{Q} \ket{\overline{\psi}_G(t)} \nonumber
\\
&= \frac{1}{2}{\rm tr}\left(\bm W^\dagger(t) \bm W(t)\right)   + [\bm \beta^*(t)]^{\rm T} \bm \beta(t).
\end{align} 
In what follows we drop the time-dependence of $\bm W(t)$ and $\bm \beta(t)$ for convenience. This quantity is conserved if its time derivative is zero
\begin{align}
\label{eq:dQ}
\frac{d \langle \hat{Q}\rangle }{dt} = 0.
\end{align}
Putting Eq. \eqref{eq:Q} into Eq. \eqref{eq:dQ} we obtain
\begin{align}
\label{eq:mainQ}
\frac{1}{2}{\rm tr}A + B=0
\end{align}
where we have defined
\begin{align}
\label{eq:Amat}
A &\equiv \frac{d \bm W^\dagger}{dt}\bm W+\bm W^\dagger\frac{d\bm W}{dt},
\\
\label{eq:Bmat}
B&\equiv \sum_{l} \left( \frac{d\beta^*_l}{dt}\beta_l + \beta^*_l\frac{d\beta_l}{dt}\right).
\end{align}
Now we use the differential equations for $\beta_l$ and $\bm W$ in Eqs. \eqref{eq:dot beta} and \eqref{eq:Wdot} to simplify $A$ and $B$. First, to simplify  $B$ we put Eq. \eqref{eq:dot beta} into Eq. \eqref{eq:Bmat}
\begin{align}
  \label{eq:Bmat1}
 B&= i  \sum_{i,k,j}\gamma_{kji}\beta_i \left[\bm V^* \bm W^\dagger \right]_{kj} -  i\sum_{i,k,j}\gamma^*_{kji}\beta^*_i \left[\bm V \bm W^{\rm T}\right]_{kj}\nonumber
 \\
 &=i\sum_{k,j}[\bm \zeta]_{kj} \left[\bm V^* \bm W^\dagger \right]_{kj}  -i\sum_{k,j}[\bm \zeta^*]_{kj}  \left[\bm V W^{\rm T}\right]_{kj} \nonumber
 \\
 &=i{\rm tr}\left(\bm \zeta \bm W^* \bm V^\dagger\right)- i{\rm tr}\left(\bm \zeta^* \bm W \bm V^{\rm T}\right).
\end{align} 
Next we simplify $A$ by putting Eq. \eqref{eq:Wdot} into Eq. \eqref{eq:Amat}, and
obtain
\begin{align}
\label{eq:Amat1}
A &=2i\bm V^{\rm T}\bm \zeta^*\bm W-2i\bm W^\dagger \bm \zeta \bm V^*.
\end{align}
Putting Eqs. \eqref{eq:Amat1} and \eqref{eq:Bmat1} into Eq. \eqref{eq:mainQ}
\begin{align}
0&=\frac{1}{2}{\rm tr}\left[  2i\bm V^{\rm T}\bm \zeta^*W-2i\bm W^\dagger \bm \zeta \bm V^*  \right]  +i{\rm tr}\left(\bm \zeta \bm W^* \bm V^\dagger\right)\nonumber
\\
&- i{\rm tr}\left(\bm \zeta^* \bm W \bm V^{\rm T}\right) \nonumber
\\
&= i{\rm tr}\left(\bm \zeta^*\bm W \bm V^{\rm T}\right)-i{\rm tr}\left(\bm \zeta \bm V^*\bm W^\dagger \right)+i{\rm tr}\left(\bm \zeta \bm W^* \bm V^\dagger\right)\nonumber
\\
&- i{\rm tr}\left(\bm \zeta^* \bm W \bm V^{\rm T}\right), \nonumber
\end{align}
and since $\bm W^*\bm V^\dagger = \bm V^*\bm W^\dagger$ the right hand side is zero.

\end{document}